\documentclass{svjour2}                    
\smartqed  
\usepackage{graphicx}
\usepackage{amssymb}
\usepackage{amsmath}
\usepackage{amssymb}
\journalname{Journal of Statistical Physics}
\begin{document}

\title{Solvable phase diagrams and ensemble inequivalence for two-dimensional and geophysical turbulent flows}

\author{Antoine Venaille         \and
        Freddy Bouchet 
}


\institute{A. Venaille \at
              GFDL-AOS Forrestal Campus Princeton NJ 08540 USA \\
              \email{venaille@princeton.edu}           
           \and
           F. Bouchet \at
             Laboratoire de Physique, Ecole Normale Sup´erieure de Lyon, Universit´e de
Lyon, CNRS, 46 All´ee d’Italie, 69364 Lyon c´edex 07, France\\
             \email{freddy.bouchet@ens-lyon.fr}
}

\date{Received: date / Accepted: date}

\maketitle

\begin{abstract}

Using explicit analytical computations, generic occurrence of inequivalence between two or more statistical ensembles is obtained for a large class of equilibrium states of two-dimensional and geophysical turbulent flows. The occurrence
of statistical ensemble inequivalence is shown to be related to previously observed phase transitions in the equilibrium flow topology. We find in these turbulent flow equilibria, two mechanisms for the appearance of ensemble equivalences, that were not observed in any physical systems before. These mechanisms are associated respectively with  second-order azeotropy (simultaneous appearance of two second-order phase transitions), and with bicritical points (bifurcation from a first-order to two second-order phase transition lines). The important roles
of domain geometry, of topography, and of a screening length scale (the Rossby radius of deformation) are discussed. It is found that decreasing the screening length scale (making interactions more local) surprisingly widens the range of parameters associated with ensemble inequivalence. These results are then generalized to a larger class of models, and applied to a complete description of an academic model for inertial oceanic circulation, the Fofonoff flow.


\end{abstract}

\section{Introduction}

One of the major interests and achievements of equilibrium statistical mechanics is the prediction of phase transitions: without describing details of the system configuration at a microscopic level, statistical mechanics predicts  drastic changes in its macroscopic properties when varying only a few key parameters. Properties and computations of these phase transitions can be very different depending on the long-rangedness or short-rangedness nature of the interaction potential characterizing the physical system. We call long-range the interaction for which the potential is non-integrable. In the case of an algebraic potential $ V(r) \sim 1/r^{\alpha}$, this occurs when $\alpha$ is less than the (spatial) dimension of the system.  
Classical examples of long-range interacting systems are two-dimensional\footnote{The interaction potential is logarithmic for two-dimensional turbulent flows.} and geophysical flows \cite{SommeriaRobert:1991_JFM_meca_Stat}, self-gravitating stars and galaxies \cite{Padmanabhan_1990,Chavanis:2002_PhysRevE_Phasetransition_SGS,IspolatovCohen:2001_PRE_PhaseTransition_SGS}, or unscreened plasmas \cite{MarcorDoveilElskens:2005_PRL_Wave_Particle}. All these systems have the property of self-organizing spontaneously into large-scale coherent structures with a simultaneous complex small-scale dynamics: self-gravitating systems form galaxies and clusters, plasmas are organized into clouds of particles and turbulent flows create vortices and jets. Taking advantage of the long-rangedness of the interactions, statistical mechanics approach is particularly suitable to study these systems: it should predict the macroscopic organization of the system as the most probable outcome of its complicated microscopic dynamics, with constraints provided by the dynamical invariants, see e.g. the Miller-Robert-Sommeria equilibrium statistical theory (MRS theory hereafter) in the context of two-dimensional and geophysical turbulent flows \cite{Robert:1990_CRAS,Miller:1990_PRL_Meca_Stat,Robert:1991_JSP_Meca_Stat,SommeriaRobert:1991_JFM_meca_Stat}.   

Phase transitions and their related thermodynamic properties are well understood and have been thoroughly studied in the context of short-range interacting systems \cite{LandauPhysStat}.  By contrast, despite the important progress in dynamics and thermodynamics of long-range interacting systems achieved over the last decade  (see e.g. \cite{EllisHavenTurkington:2000_Inequivalence,DRAW:2002_Houches,touchette04,Bouchet_Barre:2005_JSP,DRC:2008_Houches,CDR:2009_PhysRep,BouchetGuptaMukamel}) and despite many analytical examples of phase transitions in such systems (see e.g. \cite{ChavanisSommeria:1996_JFM_Classification} for two-dimensional flows,  \cite{Kiessling_Neukirch_2003_PNAS} for plasmas, \cite{Padmanabhan_1990,Stahl_Kiessling_Schindler_1995,ChavanisRevue06} for self-gravitating systems, \cite{CosteniucEllisTouchette:2005} for spin models), a clear understanding of  phase transitions in the framework of long-range interacting systems is still lacking.\\

One difficulty and challenging aspect associated with phase transitions in long-range interacting systems is the possible inequivalence between statistical ensembles: a solution in the microcanonical (constrained) ensemble is not necessarily a solution in the canonical (unconstrained) ensemble \cite{EllisHavenTurkington:2000_Inequivalence}. The physical reason for this phenomenon is the non additivity  of the energy  $E \sim \int \mathrm{d} r \ r^d V(r)$  (with $d$ the spatial dimension): when dividing a system into two subsystems, the energy of the whole system is not equal to the sum of the energy of each subsystem, because of the  non-integrability of the interaction potential $V$. Then, many classical results of thermodynamics, obtained in the framework of short-range interacting systems, and relying on the additivity of the energy, are no more valid: i) statistical ensembles are not necessarily equivalent, ii) the equilibrium entropy $S(E)$ may be non-concave. This implies that the heat capacity $C=\left( \partial T / \partial E \right)^{-1} = - T^{-2} \partial^2 S / \partial E^2$ may be negative in the microcanonical ensemble: the temperature decreases when the energy increases. Such a counter-intuitive property was first predicted in the context of astrophysics by Lyden-Bell \cite{LyndenBell:1968_MNRAS} and by Thirring \cite{Thirring70,Hertel_Thirring_1971_CMP,Hertel_Thirring_1971_AnnPhys}.

The appearance of ensemble inequivalence and negative heat capacity is associated with a zoology of phase transitions, with properties  that would not be possible in systems with short-range interactions or local interactions only (for instance, first-order phase transitions with negative jump in temperature in the microcanonical ensemble). Such peculiarities have first been revealed  in astrophysical context\cite{LyndenBell:1968_MNRAS,Thirring70,Hertel_Thirring_1971_CMP,Hertel_Thirring_1971_AnnPhys}. Further characterization of ensemble inequivalence and its associated phase transitions have then been possible by analytical studies of toy models, as for instance (among others models) simple spin systems, see e.g. \cite{BarreMukamelRuffo:2001_PRL_BEC,CosteniucEllisTouchette:2005} and references therein. Subsequently, a classification of all possible phase transitions and their link with the occurrence (or not) of ensemble inequivalence when varying an external parameter was given \cite{Bouchet_Barre:2005_JSP}. Some of the transitions predicted have not been observed yet, neither in toy models, nor in real physical systems:  this is the case for  bicritical points (bifurcation from a first order to two second-order phase transitions) and  for second-order azeotropy (simultaneous appearance of two second-order phase transitions).
Experimental observation of negative heat capacity and its related thermodynamic peculiarities remains an open problem. Given a physical system, three necessary steps to tackle this problem are i) to predict a range of parameters associated with the occurrence of ensemble inequivalence, ii) to determine the nature of associated phase transitions iii) to relate these phase transitions to qualitative changes in the systems organization.  Such is the focus of this paper, in the case of two-dimensional and turbulent geophysical flows.\\


Remarkably, any piece of progress achieved for a specific long-range interacting system is useful for another one, since the computations of statistical equilibrium states generally involve similar variational problems.
Let us for instance consider a physical system characterized by the mean field $\rho$ at a macroscopic level, with one constraint given by the energy conservation $\mathcal{E}[\rho]=E$. Assuming that the mean-field approach is valid, which is proven to be true for a large class of long-ranged interacting systems  (see e.g \cite{CagliotiLMP:1992_CMP,Eyink_Spohn_93,Kiessling_1993_PointVortex_CommPureApplMaths,kiessling_97} for the point-vortex model, \cite{Michel_Robert_1994_JSP_GRS} for two-dimensional Euler equations, and  \cite{EllisHavenTurkington:2000_Inequivalence,Barre_Bouchet_DR:2005_JSP} for a larger class of models), microcanonical and canonical problems can be described by two dual variational problems, respectively,
\begin{equation}
S(E)=\max_{\rho}\left\{ \mathcal{S}[\rho]\ | \ \mathcal{E}[\rho]=E \right\} \  \ \text{and}
\label{eq:S}
\end{equation}
\begin{equation} 
F ( \beta ) = \max_{\rho} \left \{ \mathcal{F}[\rho]=- \mathcal{S}[\rho] + \beta \mathcal{E}[\rho] \right \} \ ,
\label{eq:FF}
\end{equation}
where $\mathcal{S}$ is a Boltzmann (mixing) entropy functional and $\mathcal{F}$ a (dual) Helmholtz free energy functional. These variational problems are dual because they have the same critical points, but not the same constraints. In presence of additional constraints, the only difference is the introduction of additional statistical ensembles associated with additional dual variational problems. The validity of the mean-field approach is due the fact that the dynamics of the system at a particular location is not governed by local interactions, but rather by the whole system configuration: because every (infinitesimal) part of the system contributes to the local field, fluctuations around local mean-field values are small.\\
The aim of the present work is to present analytical computations of phase diagrams for a large class of equilibrium states of two-dimensional and geophysical turbulent flows, in the framework of the MRS theory,  emphasizing the  occurrence of ensemble inequivalence and its  related thermodynamic properties. 
The interests are fourfold.
First, beyond the application to two-dimensional and geophysical flows, the method used to compute the statistical equilibria is very general: in practice, we  solve variational problems of the form (\ref{eq:S}) and (\ref{eq:FF}), when the energy $\mathcal{E}$ and the entropy $\mathcal{S}$ are quadratic functionals, and when an additional linear constraint is taken into account\footnote{In the context of two-dimensional flows, it will be shown that this corresponds to a class of solutions characterized by a linear relation between vorticity and streamfunction.}.  Taking advantage of classical results about constrained and associated relaxed variational problems (a solution of (\ref{eq:FF}) is necessarily a solution of (\ref{eq:S}), see for instance \cite{EllisHavenTurkington:2000_Inequivalence}), the computation of statistical equilibria will amount to demanding that a quadratic functional to be positive-definite. This computation will lead to an explicit criterion for the existence of statistical equilibria in different ensembles.
Second, the computation of these equilibrium states relates previously observed phase transitions to the occurrence of ensemble inequivalence, thus deciphering  its nature and physical origin. 
On the one hand, the existence of ensemble inequivalence in the context of two-dimensional flows was proven  mathematically for point-vortices \cite{CagliotiLMP:1995_CMP_II(Inequivalence),kiessling_97} some time ago, numerically observed in quasi-geostrophic models \cite{EllisHavenTurkington:2002_Nonlinearity_Stability}, as well as in a Monte Carlo study of point vortices in a disk \cite{Smith_ONeil:1990_phys_fluid}. On the other hand, phase transitions in the flow topology of  equilibrium states have been described by Chavanis and Sommeria in the context of two-dimensional Euler flows \cite{ChavanisSommeria:1996_JFM_Classification}. The method used in the present paper to compute statistical equilibria allows  for the simultaneous description of these phase transitions and the appearance of ensemble inequivalence associated to them. In addition, we report in this context the first observation of a bicritical point and second-order azeotropy in a long-range interacting system (see also the recent works of \cite{Daux10,Stan10}), and discuss the related ensemble. 
Third, the class of models we study includes a 1.5-layer quasi-geostrophic model, which allows to discuss changes of phase diagrams in the presence of a screening length scale (Rossby radius of deformation). When this parameter tends to zero, the energy is additive at leading order. However, it will be shown that non-local first-order corrections of the energy are essential to the computation of the equilibrium states, thereby  explaining  the apparent paradox that decreasing the screening length scale widens the ensemble inequivalence area in the phase diagram. 
Fourth, a generalization of these results allows to describe a variety of ``Fofonoff modes'' characterized by a flow structure differing from the classical low-energy solution of inertial academic ocean models \cite{Fofonoff:1954_steady_flow_frictionless}. We provide a complete description of these modes,  and explain why high-energy states could not have been obtained in previous computations of Fofonoff flows, for they did not take into account the possible existence of ensemble inequivalence. 
Some of these results have been already presented in a letter \cite{Venaille_Bouchet_PRL_2009}. 
We present in this longer version detailed computations, discussions, and generalizations of this previous work.\\

The paper is organized as follows: In section 2 we review a class of models for two-dimensional and geophysical turbulent flows, as well as the statistical mechanics of these systems. In section 3 we recall classical results on the equivalence between statistical ensembles, and explain how they can be used efficiently to compute equilibrium  states of two-dimensional and geophysical flows; we then apply these methods to explicit calculations of the phase diagrams associated with any flow characterized by a linear relation between potential vorticity and stream function ($q-\psi$ relation). In section 4 we present the thermodynamic properties of these systems first in the case of the Euler dynamics for two-dimensional turbulent flows; we report the generic existence of ensemble inequivalence, and describe phase transitions in the flow structure associated with it, including a bicritical point. In section 5 we discuss generalization of these computations to a wider class of models and apply these results to a complete description of Fofonoff flows. This generalization makes it possible to describe second-order azeotropy. 

\section{Equilibrium states of two-dimensional flows}

The aim of this section is to introduce a model for inviscid two-dimensional and geophysical turbulence (Euler equations and quasi-geostrophic equations), and to review the equilibrium statistical mechanics of these systems.

\subsection{Models: Euler and quasi-geostrophic equations}

We consider in this paper a class of models that can be expressed as the transport of a scalar quantity  $q(\mathbf{r})$ by an incompressible, two-dimensional velocity field, where $\mathbf{r}=(x,y)$ is the spatial coordinate: 
\begin{equation}
\partial_{t}q+\boldsymbol{u\cdot\nabla}q=0 \quad \text{with} \quad  \mathbf{u}=\left(-\frac{\partial \psi}{\partial y}\ ,\ \frac{\partial\psi}{\partial x}\right)\quad \text{and} \quad q=\Delta\psi-\frac{\psi}{R^{2}}+h \ , \label{eq:qg_model_0_bord}
\end{equation}
where $\psi(\mathbf{r})$ is a streamfunction, $h(\mathbf{r})$ a prescribed two-dimensional field interpreted as a topography (in addition, and without loss of generality, it will be assumed in the following that  $\int_{\mathcal{D}} \mathrm{d} \mathbf{r} \  h =0$),  and $R$ the Rossby radius of deformation, an intrinsic length scale of the system.
One must in addition specify a boundary condition for $\psi$: we consider in this paper the case of a closed domain $\mathcal{D}$, with impermeability constraint: there is no flow across the boundary $\partial \mathcal{D}$.
This means that the streamfunction is constant along the boundary. 
We consider here the simplest case $\psi=0$.
Generalization to different ways of specifying the constant streamfunction at the boundary\footnote{In geophysical context, it is more physically relevant to consider $\psi=\psi_{fr}$ at boundaries, with $\psi_{fr}$ is determined by a mass conservation constraint reading $\int_\mathcal{D} \mathrm{d} \mathrm{r} \ \psi =0$. The result presented in this paper would not be different in that case, but this lead to slightly more complicated computations.}, or even to other domain geometries (channel, doubly periodic)
is straightforward. 
When $R=+\infty$ and $h=0$, $q$ is the vorticity and equations (\ref{eq:qg_model_0_bord}) are the two-dimensional Euler equations \cite{euler250}.
When $R=+\infty$ with possibly $h \ne 0$, this is the barotropic quasi-geostrophic dynamics in presence of topography, which has been thoroughly studied in oceanic or atmospheric context \cite{PedloskyGFD}, and in which case $q$ is the potential vorticity.
When $R$ is finite, this is the $1.5$ layer quasi-geostrophic model (also called equivalent barotropic), a refinement of the barotropic model which includes the effect of stratification.
This model is also relevant to describe plasma turbulence, in which case $R$ is interpreted as the Larmor radius.
When the topography is $h = b y$, this is the Charney-Hazegawa-Mima model for either  geophysical quasi-geostrophic turbulence on a beta plane or drift plasma turbulence \cite{Charney49,HazegawaMima78}. 

%

\subsubsection*{Conserved quantities.} 

According to Noether's theorem, there is a dynamical invariant associated with each symmetry of the system, see e.g. \cite{Salmon_1998_Book} for a detailed discussion in the framework of geophysical flows. 
These dynamical invariants  provide important constraints for the admissible states. 

The constraint associated with time invariance is the conservation  of the total energy: 

\begin{equation}
\mathcal{E}[q]=\frac{1}{2}\left\langle \left(\nabla\psi\right)^{2}+\frac{\psi^{2}}{R^{2}}\right\rangle =-\frac{1}{2}\left\langle \left(q-h\right)\psi\right\rangle \ , \label{eq:EE}
\end{equation}
where $\left\langle \cdot \right\rangle$ stands for a spatial integration over the whole domain $\mathcal{D}$, of area $\left|\mathcal{D}\right|=1$, and where the last equality is obtained by performing an integration by parts. 
Because the dynamics is the transport of potential vorticity $q$ by an incompressible flow, it conserves any  Casimir functional  
\[
\mathcal{C}_{g}[q]=\left\langle g(q)\right\rangle \ ,\] 
 where $g$ is any continuous function on $\mathcal{D}$. 
These conservation laws are equivalent to the conservation of the global distribution of potential vorticity, and are related to the particle relabelling symmetry \cite{Ripa81}.
In the present study, a Casimir functional that will be of particular interest (because of its simple, linear form), is the integral of the potential vorticity over the whole domain:
\begin{equation}
\mathcal{C}[q]=\left\langle q\right\rangle \ .\label{eq:CC}
\end{equation}
If the domain has particular additional symmetries, for instance translational invariance (case of a channel), or rotational invariance (case of a circular geometry), then additional conserved quantities have to be taken into account.
We do not consider these situations in the present paper.

\subsubsection*{Long-range interactions}
The energy (\ref{eq:EE}) is the sum of a kinetic term $\left(\frac{1}{2}\left\langle \left(\nabla\psi\right)^{2}\right\rangle\right)$ and a gravitational potential term $\left(\frac{1}{2}\left\langle \frac{\psi^{2}}{R^{2}}\right\rangle\right)$, but it can formally be expressed as a single potential energy term
\begin{equation} 
\mathcal{E}[q]=-\frac{1}{2}\left\langle \left(q-h\right)\psi\right\rangle =-\frac{1}{2}\int\left(q(\mathbf{r})-h(\mathbf{r})\right)\mathcal{G}(\mathbf{r},\mathbf{r}^{\prime})\left(q(\mathbf{r}^{\prime})-h(\mathbf{r}^{\prime})\right)d\mathbf{r}d\mathbf{r}^{\prime} \ ,
\label{eq:longrange}
\end{equation}

where $\mathcal{G} \left( \mathbf{r},\mathbf{r}^{\prime} \right)$ is the Green function solution of 

\begin{equation} 
\Delta\mathcal{G}-\frac{\mathcal{G}}{R^{2}}=\delta(\mathbf{r}-\mathbf{r}^{\prime})\quad\text{with}\quad\mathcal{G}=0\quad\text{for }\mathbf{r}\text{ on}\quad\partial\mathcal{D} \ . 
\label{eq:green}
\end{equation}
Far from boundaries, the green function is $\mathcal{G}=\frac{1}{2\pi}K_{0}\left(|\mathbf{r}-\mathbf{r}^{\prime}|/R\right)$ where $K_{0}$ is the modified Bessel function of order zero. 
At short distance ($|\mathbf{r}-\mathbf{r}^{\prime}|\ll R$), the potential vorticity reads $q \approx \Delta \psi +h$, and the dynamics is that of the 2D Euler equations (or of the barotropic model if $h\ne0$), so the interaction potential has the classical logarithmic form: $K_{0}\sim-\ln|\mathbf{r}-\mathbf{r}^{\prime}|/R$. 
At large distance ($|\mathbf{r}-\mathbf{r}^{\prime}|\gg R$), the interaction potential becomes $K_{0}\sim\left(\frac{\pi}{2|\mathbf{r}-\mathbf{r}^{\prime}|}\right)^{1/2}e^{-|\mathbf{r}-\mathbf{r}^{\prime}|/R}$ and is screened at scale $R$. 
If the screening parameter is much smaller than the typical length scale $L\equiv \sqrt{\left|\mathcal{D}\right|}$ of the domain where the flow take place  ($R \ll L$), then the system is short-ranged. 
Otherwise the system is long-range (for $R=O(L)$ or $R \gg L$ ). The screening length scale $R$ is thus a parameter that allows one to pass from a short-range to a long-range interacting systems. 
%
%


\subsection{Statistical mechanics of two-dimensional and geophysical turbulent flows}

The class of models we study are known to develop complex vorticity filaments at finer and finer scales.  This makes almost impossible any attempt to have a deterministic approach of these systems, i.e. to describe the temporal evolution of the fine grained structure of the flow. Rather than describing these fine-grained structures, equilibrium statistical theories, assuming ergodicity, predict the final organization of the flow on a coarse grained level \cite{SommeriaRobert:1991_JFM_meca_Stat,RobertSommeria:1992_PRL_Relaxation_Meca_Stat,Miller:1990_PRL_Meca_Stat}.
To keep track of the conservation laws of the transport equation,
one needs to introduce the probability density $\rho(\mathbf{r},\sigma)$
of finding the potential vorticity level $\sigma$ at point $\mathbf{r}=(x,y)$, with a local normalization constraint 
\[N[\rho] = \int_{\Sigma} \mathrm{d} \sigma \ \rho \sigma = 1\ ,\]
where $\Sigma$ stands for the admissible set of potential vorticity levels. 
The conservation of the global vorticity distribution of potential vorticity is expressed as  
\[D \left[ \rho \right] \left(\sigma \right) = \left< \ \rho \right>\ .\]
Finally, another constraint is provided by the mean-field energy 
\[\mathcal{E} \left[ \rho \right] = \left<  \ \int_{\Sigma}  \mathrm{d} \sigma \ \left( \sigma -h\right) \rho \overline{\psi} \right> \ ,\]
where $\overline{\psi}$ is the solution of 
\[\overline{q}-h=\int_{\Sigma} \mathrm{d} \sigma \ \sigma \rho - h= \Delta \overline{\psi} -\frac{\overline{\psi}}{R^2} \ .\]
The energy of the microscopic potential vorticity fluctuations is negligible with respect to the mean-field energy of the statistical equilibrium state \cite{SommeriaRobert:1991_JFM_meca_Stat}, which can be rigorously justified using large deviation theory \cite{Michel_Robert_LargeDeviations1994CMaPh.159..195M}.
A \emph{macroscopic} or \emph{coarse-grained} state is then fully described by   the probability density $\rho$. 
The \emph{microscopic} states are the \emph{fine-grained} potential vorticity field of the flow. 
The number of \emph{microscopic} states (any fine-grained field $q$) associated with a given \emph{macroscopic} state (the probability density field $\rho$) can be quantified by the Boltzmann-Gibbs mixing entropy 

\[\mathcal{S}_{MRS}[\rho]=- \left<\int_{\Sigma} \mathrm{d} \sigma \ \rho\ln\rho \right> \ . \]
This has been rigorously justified using large deviations tools \cite{Michel_Robert_1994_JSP_GRS,EllisHavenTurkington:2000_Inequivalence}. Large deviation theory allows in addition to prove that an overwhelming number of microstates is associated with the most probable macroscopic state, i.e. the one that maximizes the mixing entropy while satisfying the constraints of the problem.
To conclude, MRS statistical equilibria are solutions of the variational problem%
\begin{equation}
S_{MRS}\left(E,d\left(\sigma\right)\right)=\max_{\left\{ \rho \right\} }\left\{ \mathcal{S}_{MRS}[\rho]\ |\ \mathcal{E}[\rho]=E\ \&\ D_{\sigma}[\rho]=d(\sigma) \  \& \ N[\rho]=1 \right\} \ ,
\label{eq:rsm_var_prob}
\end{equation}
where $E$ and $d(\sigma)$ are prescribed values of energy and  given distribution of potential vorticity levels. $S_{MRS}\left(E,d\left(\sigma\right) \right)$ is the equilibrium entropy. 
Because this variational problem takes into account all the invariants of the dynamics, we call it \emph{full microcanonical} variational problem\footnote{The term \emph{microcanonical} will be used in the following when energy and circulation are the only constraints taken into account}.
Computing the solutions of this variational problem (\ref{eq:rsm_var_prob})
 requires the knowledge of the infinity of Casimir constraints (or equivalently the knowledge of the distribution $d(\sigma)$).
This is a huge practical limitation to apply this theory, for two reasons. 
First, one does generally not know precisely what is the initial condition. 
Second, and more importantly, a variational problem with an infinite number of constraints is very difficult to handle.  
Many efforts have recently been devoted to alternative approaches which could lead to practical and mathematical simplification of this problem, see e.g. \cite{EllisHavenTurkington:2000_Inequivalence,Bouchet:2008_Physica_D}.
We present one of these methods in the following, and apply it to the actual computation of MRS equilibrium states. 

\subsection{Simplification of the variational problem of the MRS theory}

Following the idea that less constrained, dual variational problems are easier to solve than more constrained ones (see e.g. \cite{EllisHavenTurkington:2000_Inequivalence}),  it has been proposed to treat the Casimir invariants canonically to compute MRS
equilibria \cite{Bouchet:2008_Physica_D}, by studying the following variational
problem:

\begin{equation}
S(E,\Gamma)=\max_{q}\left\{ \mathcal{S}[q]\ =\left\langle s(q)\right\rangle |\ \mathcal{E}[q]=E\ \&\ \mathcal{C}[q]=\Gamma\right\} \ .\label{eq:microcanonical_var_prob}
\end{equation}
 where $s(q)$ is a concave function. Any solution to (\ref{eq:microcanonical_var_prob}) is a MRS equilibrium, i.e. a solution of  (\ref{eq:rsm_var_prob}), but the converse may be wrong. A proof and a wider explanation of this statement is developed in \cite{Bouchet:2008_Physica_D}. The important point is that considering the  coarse-grained potential vorticity fields $q$ as an order parameter rather  than the full distribution function (and treating canonically  the constraints on the potential vorticity distribution)  simplify the solution of the variational problem, not modifying the structure of critical points.

We call this problem \emph{microcanonical} by analogy with usual problems
in thermodynamics, involving a constraint on the energy, and on the
number of particle (here the circulation) in the maximization of a functional. Note that since a maximizer $q_{MC}$ of (\ref{eq:microcanonical_var_prob}) is  a coarse-grained potential vorticity field, saying that this solution is a MRS statistical equilibria means that there exists at least a set of constraints (the energy and the global vorticity distribution) for which the MRS equilibrium states $\rho_{MRS}$ satisfy $\int_\Sigma \mathrm{d} \sigma \ \rho_{MRS} =q_{MC}$.\\


The full microcanonical ensemble is relevant to describe an isolated system, in which the infinite number of constraints is fixed. 
In practice, this applies to situations where the inertial time scale of turbulent organization of the flow is much smaller than the dissipation and forcing time scales. 
We emphasize that considering a simpler variational problem, in a dual, less constrained statistical ensemble has no other reason that simplifying the mathematics.  
We interpret all the states that we compute as MRS equilibrium state in the \emph{full microcanonical ensemble}.
A physical meaning of the other, less constrained ensembles would require the introduction of a bath of potential vorticity distribution, which would lead to fuzzy discussions, and has no reasonable physical justifications. 
Similarly, because the Casimir functional $\mathcal{S}[q]$ is maximized in the microcanonical problem (\ref{eq:microcanonical_var_prob}), it will be referred to as an entropy in the following, but only the Boltzmann-Gibbs (or mixing) entropy  functional $\mathcal{S}_{MRS}$ of the variational problem (\ref{eq:rsm_var_prob}) has the physical meaning of an entropy.
%


\subsection{Critical points of the MRS variational problem}

\subsubsection{Statistical equilibria are dynamical equilibria}

%
Critical points of the microcanonical variational problem satisfy $\delta\mathcal{S}-\beta\delta\mathcal{E}-\gamma\delta\Gamma=0$, where the first variations are taken with respect to the potential vorticity field $q$, and where  $\beta$ and $\gamma$ are the Lagrange multipliers associated with the energy and the circulation conservation respectively. This yields 

\begin{equation}
q=s^{\prime-1}\left(-\beta\psi+\gamma\right)\ , \label{eq:ff}
\end{equation}
where $s^{\prime-1}$ is the inverse of the derivative of $s$, and  where the inverse temperature  $\beta$ and the fugacity $\gamma$ are prescribed by the constraints on $E$ and $\Gamma$.
We conclude that  critical points of (\ref{eq:microcanonical_var_prob})  are flows characterized by  a  $q-\psi$ relation directly related to the function $s$. 
 Such flows satisfy $\mathbf{u} \cdot \nabla q =0$ and are therefore dynamical equilibria of  the transport equation (\ref{eq:qg_model_0_bord}).

\subsubsection{The case of the maximization of a quadratic functional}

In the following, we shall compute solutions of the microcanonical problem (\ref{eq:microcanonical_var_prob}) in the case of a quadratic functional by considering  $s(q)=-\frac{1}{2}q^{2}$:
\begin{equation}
\mathcal{S}[q]=-\frac{1}{2}\left\langle q^{2}\right\rangle \ ,\label{eq:SS}
\end{equation}
The functional $\mathcal{S}[q]$ is the opposite of the enstrophy in that case. 

There are two reasons for studying this particular functional. First, it is a class of genuine statistical equilibria, with the property  of being technically very simple. This example allows for a direct computation of the variational problem (\ref{eq:microcanonical_var_prob}) and gives analytic examples of ensemble inequivalence. Second, whatever the function  $s(q)$ is, computation of statistical equilibria in the limit of low-energy require studying the quadratic problem (\ref{eq:SS}).
Let us be more precise, considering the simple case when $h=0$. The energy $E$ is then a positive-definite functional of $q$. The limit of vanishing energy implies then that $q$ is small. A multiplication of $s(q)$ by a  positive constant and the addition of a term linear in $q$ does not change solutions of (\ref{eq:microcanonical_var_prob}). Assuming $s$ twice differentiable, one can therefore consider without loss of generality that  $s(0)=0$,  $s^{\prime}(0)= 0$, and $s^{\prime\prime} (0)=-1$. A Taylor expansion gives then
\begin{equation}
s(q)= -\frac{1}{2}q^2 +O(q^{3}) \ . \label{eq:taylor}
\end{equation} 

In the case of a purely quadratic functional, using equation (\ref{eq:ff}),  we find that the corresponding $q-\psi$ relation is linear: 
\[ q = \beta\psi-\gamma\ .\] This linear relation depends on two parameters, $\beta$ and $\gamma$. These two parameters have to be computed by considering the two independent constraints of the problem (\ref{eq:microcanonical_var_prob}), namely the energy $E$ and the circulation $\Gamma$. 

In the context of the Euler equations ($R=+\infty,\ h=0$), Chavanis
and Sommeria have computed and classified MRS equilibrium states corresponding
to a linear $q-\psi$ relation, in an arbitrary closed domain \cite{ChavanisSommeria:1996_JFM_Classification}. They found two classes of phase diagrams, depending on the domain geometry. When the domain has a symmetry axis and is sufficiently stretched in the direction perpendicular to this axis, a transition from a monopole to a dipole  is observed when varying the parameter $E/\Gamma^2$ above a critical threshold. When the domain is not stretched enough along the direction perpendicular to the symmetry axis, no such transition is observed when varying the parameter $E/\Gamma^2$.
In \cite{ChavanisSommeria:1996_JFM_Classification}, the method to compute the MRS equilibrium states was, first to compute all the stationary solutions associated with a linear $q-\psi$ relation, and then to compute their entropy in order to select the entropy maxima. 
This made possible the observation (through analytical calculations) of the intriguing phase transitions described above, but the precise nature of these transitions is still unknown, as well as the underlying physical mechanism responsible for it. 
We show in the following that the use of another analytical method, taking advantage of the aforementioned relation between constrained and dual, less constrained variational problems, allows for the first time i) to compute explicitly statistical ensemble inequivalence in two-dimensional turbulent flows, ii) to relate this statistical ensemble inequivalence to the occurrence of the phase transitions mentioned in previous paragraphs iii) to generalize these results to a larger class of models.

\section{Computation of MRS equilibrium states and ensemble
inequivalence}

In this section we solve the variational problem (\ref{eq:microcanonical_var_prob}) when $S[q]=-\left\langle q^2 \right \rangle/2 $, and classify the solutions  according to their energy and circulation.

\subsection{General method \label{sub:method}}

The strategy to solve (\ref{eq:microcanonical_var_prob}) is to introduce and solve dual variational problems for which one or two of the constraints are relaxed.
The natural first step is to consider the less constrained problem, referred to as \emph{grand-canonical}, which is the easiest to solve:
\begin{equation}
J(\beta,\gamma)=\min_{q}\left\{ \mathcal{J}[q]=-\mathcal{S}[q]\ +\beta\ \mathcal{E}[q]+\gamma\mathcal{C}\left[q\right]\right\} \ ,\label{eq:grandcanonical}
\end{equation}
where $J$ is a thermodynamic potential. We compute its solutions, their energy and circulation.  We then check which part of the plane  $E,\Gamma$ is filled by these solutions. Then we use the central result that any solution $q_{gc}$ of the relaxed variational problem (\ref{eq:grandcanonical}) is  also a solution of the more constrained variational problem (\ref{eq:microcanonical_var_prob}) with constraints given by $E=\mathcal{E}[{q_c}]$ and $\Gamma=\mathcal{C}[q_{gc}]$. If  some range of admissible parameters $E,\Gamma$ are not achieved by grand-canonical solutions, one must consider a more constrained ensemble, and so on. In our case, it will be sufficient to consider the problem with only one constraint on the circulation, referred to as the \emph{canonical} variational problem, in order to find all the solutions on the plane  $E,\Gamma$: 
\begin{equation}
F(\beta,\Gamma)=\min_{q}\left\{ \mathcal{F}[q]=-\mathcal{S}[q]\ +\beta\ \mathcal{E}[q]\ |\ \mathcal{C}[q]=\Gamma\right\} \ ,\label{eq:canonical}
\end{equation}
where $F$ is a free energy. The fact that any value of the parameter ($E,\Gamma$) will be achieved by a canonical solution  proves the equivalence between this ensemble and the microcanonical one. We will show in addition the existence of a range of parameter $E,\Gamma$ for which grand-canonical and canonical problems are not equivalent: for this range of parameters, there is a canonical solutions, but no grand-canonical solutions.  
In the present case, the functional $\mathcal{E}$ and $\mathcal{S}$ are quadratic and the functional $\mathcal{C}$ is linear. It is possible to take advantage of the linearity of this constraint in order to express the canonical problem (\ref{eq:canonical}) in the form of an unconstrained variational problem involving the minimization of a quadratic constraint. Then both problem (\ref{eq:grandcanonical}) and (\ref{eq:canonical}) are solved by finding the minimum of a quadratic functional,  with a possible linear part and no constraints. Let $Q$ and $L$ be the linear operators associated to the purely quadratic part and to the purely linear part of this functional respectively. Then we have three cases:
\begin{itemize}
\item \textbf{Case 1.} The smallest eigenvalue of $Q$ is positive: the
minimum exists and is achieved by an unique minimizer. 
\item \textbf{Case 2.} At least one eigenvalue of $Q$ is strictly negative.
There is no minimum. 
\item \textbf{Case 3.} The smallest eigenvalue of $Q$ is zero (with eigenmode
$e_{neutral}$). If $L[e_{neutral}]=0$ (case \textbf{3a}), the minimum exists,
and each state of the neutral direction $\left\{ \alpha e_{neutral}\right\}_{\alpha \in \mathbb{R}} $
is a minimizer. If $L[e_{neutral}]\ne0$ (case \textbf{3b}), then no minimum
exists.
\end{itemize}
We see that in the present case,  the computation of statistical equilibria involves only the diagonalization of a quadratic operator. 
%
%
Application of this general method and detailed computations are provided in the remaining of this section.

\subsection{Decomposition on Laplacian eigenmodes \label{sub:Laplacian}}

In order to compute eigenvalues of the quadratic operators involved in the grand-canonical problem (\ref{eq:grandcanonical}) and in  the canonical problem (\ref{eq:canonical}), it is convenient to project the different fields ($q$, $h$
and $\psi$) on Laplacian eigenmodes.
%
We introduce the complete, orthonormal basis $\left\{ e_{i}(\mathbf{r})\right\} _{i\in\mathbb{N}}$
of Laplacian eigenmodes on the domain $\mathcal{D}$ 
\begin{equation}
\left(\Delta-R^{-2}\right)e_{i}=-\mu_{i}e_{i}\ =-\left(\lambda_{i}+R^{-2}\right)e_{i}\ . \label{eq:def_mu}
\end{equation}
 The Laplacian eigenvalues $\lambda_{i}$ are all positive, in increasing
order. 
Potential vorticity $q$ and topography $h$ can be decomposed on this basis,
as well as the streamfunction, computed by inverting the relation
$q-h=\left(\Delta-R^{-2}\right)\psi$: 
\[
q=\sum_{i}q_{i}e_{i}\quad\quad h=\sum_{i}h_{i}e_{i}\quad\quad\psi=\sum_{i}\frac{h_{i}-q_{i}}{\mu_{i}}e_{i}\ .\]
 The functionals $\mathcal{S}$, $\mathcal{E}$
and $\mathcal{C}$ given by equations (\ref{eq:SS}), (\ref{eq:EE}),
and (\ref{eq:CC}) respectively  can then be expressed
in terms of the coordinates $\left\{ q_{i}\right\} $:

\begin{eqnarray}
 & \text{} & \mathcal{S}[q]=-\frac{1}{2}\sum_{i}q_{i}^{2}\ ,\label{eq:S_qi}\\
 & \text{} & \mathcal{E}[q]=\frac{1}{2}\sum_{i}\frac{1}{\mu_{i}}\left(q_{i}-h_{i}\right)^{2}\ ,\label{eq:E_qi}\\
 & \text{} & \mathcal{C}[q]=\sum_{i}q_{i}\left\langle e_{i}\right\rangle \ ,\label{eq:C_qi}\end{eqnarray}

where $\left\langle e_{i}\right\rangle =\int_{\mathcal{D}} \mathrm{d} \mathbf{r} \ e_{i}(\mathbf{r})$
.\\

We introduce two independent subspaces: 
\begin{itemize}
\item the subspace of the Laplacian
eigenmodes having zero mean value ( $\left\langle e_{i}^{\prime}\right\rangle =0$,
$i\in\mathbb{N}^{*}$) and
\item the subspace of the Laplacian eigenmodes
having non-zero mean value ( $\left\langle e_{i}^{\prime\prime}\right\rangle \ne0$,
$i\in\mathbb{N}^{*}$). 
\end{itemize}
The notation {}``prime'' and {}``double
prime'' will be used to distinguish (when necessary) one subspace
from the other.\\

The subspace of zero mean Laplacian eigenmode
$\left\{ e_{i}^{\prime}\right\} $ is generically empty when the
domain geometry $\mathcal{D}$ admits no particular symmetry: a small
perturbation of the domain would change the mean value of the eigenmode. If, by contrast, one imposes a symmetry axis in the domain geometry,
then it exits generically eigenmodes having a zero mean: these eigenmodes
are antisymmetric with respect to the symmetry axis.  This is for
instance the case for a rectangular domain, for which the eigenmodes
and eigenvalues are given in Annexe C.

We assume that the smallest Laplacian eigenvalue on a closed domain is not degenerate, and that the corresponding eigenmode is positive everywhere, which turn out to be the case for simply connected bounded domains. It is thus an eigenmode of non-zero mean value: $e_{1}=e_{1}^{\prime\prime}$, and $\mu_{1}=\mu_{1}^{\prime\prime}$. Then, we have necessarily $\mu_{1}^{\prime\prime}<\mu_{1}^{\prime}$.

\subsection{Solution for the grand-canonical problem \label{subsec:solution_grandcanonical}}

We consider the grand-canonical problem (\ref{eq:grandcanonical}). Using expressions (\ref{eq:S_qi}), (\ref{eq:E_qi}) and (\ref{eq:C_qi}) for entropy, energy and circulation functionals, we find:
\begin{equation}
\mathcal{J}[q]=\frac{1}{2}\sum_{i\ge1}\left(1+\frac{\beta}{\mu_{i}}\right)q_{i}^{2}+\sum_{i\ge1}\left(-\frac{\beta}{\mu_{i}}h_{i}+\gamma\langle e_{i}\rangle\right)q_{i}\ .\label{eq:J}
\end{equation}
Since the quadratic part of $\mathcal{J}$ is diagonal in the Laplacian
eigenmode basis, we can immediately see that there is a unique solution
to the variational problem if and only if $\beta>-\mu_{1}$ (case
1. of subsection \ref{sub:method}), whatever the value of $\gamma$. 
If $\beta<-\mu_{1}$
(case 2. of subsection \ref{sub:method}), there is no solution to the variational problem.
If $\beta=-\mu_{1}$, then $Q[e_{1}]=0$ (case 3. of subsection \ref{sub:method}) and there
is a neutral direction if and only if $L[e_{1}]=0$, which yields
$\gamma=h_{1}/\left\langle e_{1}\right\rangle $.\\

We show in Appendix \ref{app:solution_grandcanonical} that 
 each point in the plane $(E,\Gamma)$ located
on the set of parabolas $\left\{ E_{\beta}(\Gamma)\ |\ \beta>-\mu_{1}\right\} $
corresponds to a unique grand-canonical solution, where $E_{\beta}(\Gamma)$
is given by

\begin{equation}
E_{\beta}(\Gamma)=\mathcal{A}_{\beta}[h]+\mathcal{B_{\beta}}[h]\Gamma+\left(\frac{1}{2\left(f(\beta)\right)^{2}}\sum_{i\ge1}\frac{\mu_{i}\langle e_{i}\rangle^{2}}{\left(\mu_{i}+\beta\right)^{2}}\right)\Gamma^{2}\ ,\label{eq:E_G}\end{equation}

(see (\ref{eq:E_G_app})), and where $\mathcal{A}_{\beta}[h]=\mathcal{B}_{\beta}[h]=0$ for $h=0$, and with 
\begin{equation}
f(\beta)=\sum_{i\ge1}\frac{\mu_{i}\langle e_{i}\rangle^{2}}{\mu_{i}+\beta}\ .\label{eq:f_beta} 
\end{equation}

Interestingly, The curvature of the parabola $E_{\beta}(\Gamma)$ equation (\ref{eq:E_G}) does not depend on the topography $h$, and is a decreasing function of
$\beta$. 
Notice also that the lower bound for the (admissible) energies among all flows with a given circulation is achieved when the inverse temperature tend to infinity $\beta \rightarrow + \infty$. By taking this limit in equations (\ref{eq:E_G}), (\ref{eq:f_beta}) and using $\left<h\right>=0$ (see section $2.1$), we find that these energy minima are located on the parabola 

\begin{equation}
E_{m}(\Gamma)={\mathcal{B}_{m}}[h]\Gamma+\frac{1}{2}\left(\sum_{i\ge1}\mu_{i}\langle e_{i}\rangle^{2}\right)^{-1}\Gamma^{2}\ ,\label{eq:energy_infty}
\end{equation}
where $\mathcal{B}_{m}[h]=0$ for $h=0$. 
We conclude that grand-canonical solutions cover the whole area above
the parabola $E_{m}(\Gamma)\ $, given by (\ref{eq:energy_infty})
and below the parabola $E_{-\mu_{1}}(\Gamma)$ obtained by taking $\beta\rightarrow -\mu_1 $ in (\ref{eq:E_G}) (see also equation (\ref{eq:parabola_inequivalence}) in Appendix \ref{app:solution_grandcanonical}).
Energies located above the parabola $E_{-\mu_{1}}(\Gamma)$
are not achieved by grand-canonical solutions, and the values located
below $E_{m}(\Gamma)$ are not admissible. Because it exists
a range of admissible energies and circulations not reached by grand-canonical solutions, we are in a situation of ensemble inequivalence. We then turn to the more constrained canonical problem to find solutions
in this area.

\subsection{Solution for the canonical problem}

We now consider the canonical problem (\ref{eq:canonical}). It is possible to transform this constrained problem into an unconstrained
variational problem, taking advantage of the linearity of the circulation constraint. Using this constraint, and recalling that $\left<e_{1}\right>\ne0$, one coordinate can be expressed in terms of the others: 

\begin{equation} 
\widetilde{q} \equiv q-q_1 e_1 (\mathbf{r}) \quad \text{with} \quad q_{1}= \frac{\Gamma-\sum_{i\ge2}q_{i}\left\langle e_{i}\right\rangle}{\left\langle e_{1}\right\rangle} \label{eq:qtilde} 
\end{equation} 

This expression is injected into the functional $\mathcal{F}=-\mathcal{S}+\beta\mathcal{E}$ appearing in the variational problem (\ref{eq:canonical}):

\begin{eqnarray}
\mathcal{F}[\widetilde{q}] & = & \frac{1}{2}\sum_{i\ge2}\left(1+\frac{\beta}{\mu_{i}^{\prime}}\right){q_{i}^{\prime}}^{2}-\sum_{i\ge2}\frac{\beta}{\mu_{i}^{\prime}}h_{i}^{\prime}q_{i}^{\prime}\nonumber \\
 &  & +\frac{1}{2}\sum_{i,j\ge2}\left(\delta_{ij}\left(1+\frac{\beta}{\mu_{i}^{\prime\prime}}\right)+\left(1+\frac{\beta}{\mu_{1}^{\prime\prime}}\right)\frac{\langle e_{i}^{\prime\prime}\rangle\langle e_{j}^{\prime\prime}\rangle}{\langle e_{1}^{\prime\prime}\rangle^{2}}\right)q_{i}^{\prime\prime}q_{j}^{\prime\prime}\nonumber \\
 &  & -\sum_{i\ge2}\left(\Gamma\frac{\langle e_{i}^{\prime\prime}\rangle}{\langle e_{1}^{\prime\prime}\rangle{}^{2}}\left(1+\frac{\beta}{\mu_{1}^{\prime\prime}}\right)-\frac{\beta}{\mu_{i}^{\prime\prime}}h_{i}^{\prime\prime}+\frac{\langle e_{i}^{\prime\prime}\rangle}{\langle e_{1}^{\prime\prime}\rangle}\frac{\beta}{\mu_{1}^{\prime\prime}}h_{1}^{\prime\prime}\right)q_{i}^{\prime\prime}\ ,\label{eq:F}\end{eqnarray}
where we have made the distinction made in subsection \ref{sub:Laplacian} between the two (independent) subspaces of Laplacian eigenmodes. The problem is now to find the minimizer $\widetilde{q}$ of this functional, with no constraint.\\

To find the solutions to this problem, we use the same method as in
the grand-canonical case. We call $Q$ and $L$ the linear operators
associated with the purely quadratic and linear parts of $\mathcal{F}$.
In the subspace of zero-mean Laplacian eigenmodes, $Q$ is diagonal.
Its smallest eigenvalue is strictly positive if and only if $\beta>-\mu_{1}^{\prime}$, where $\mu_{1}^{\prime}$ is the smallest eigenvalue of the linear
operator $-(\Delta-R^{-2})$ associated with an eigenmode with zero
mean-value.

In the subspace of non zero-mean Laplacian eigenmodes, $Q$ is not
diagonal, so this case requires more computations. We look for the
value $\beta=-\mu^*$ such that the smallest eigenvalue of $Q$ is zero
in this subspace (that corresponds to case 3. of subsection \ref{sub:method}). 
We thus look for the eigenmode  

\[\widetilde{e}^{*}=\sum_{i\ge2} e_{i}^{*}e_{i} \quad \text{such that } Q[\widetilde{e}^{*}]=0 \ ,\]
which yields:
\begin{equation}
\forall i\ge2,\ \left(1-\frac{\mu^{*}}{\mu_{i}}\right)e_{\ i}^{*}+\frac{\langle e_{i}\rangle}{\langle e_{1}\rangle^{2}}\left(1-\frac{\mu^{*}}{\mu_{1}}\right)\sum_{j\ge2}\langle e_{j}\rangle e_{j}^{*}=0 \ .\label{eq:vp_q_star_intermediaire}\end{equation}

Let us first assume $\sum_{j\ge2}\langle e_{j}\rangle e_{j}^{*} = 0$. Equation (\ref{eq:vp_q_star_intermediaire}) then implies that there exists an integer $k>2$ such that $\mu^*=\mu_k$, with $e_i^*=0$  for $i \ne k$, and  $e_k^*=\alpha$, where $\alpha$ a real. But then we have $\sum_{j\ge2}\langle e_{j}\rangle e_{j}^{*} = \alpha$, which contradicts the initial hypothesis if $\alpha \ne 0$.  We conclude that $\sum_{j\ge2}\langle e_{j}\rangle e_{j}^{*}\ne0$.
Using this result, and multiplying (\ref{eq:vp_q_star_intermediaire})
by $\mu_i\langle e_{i}\rangle/\left(\mu_i-\mu^{*}\right)$ 
yields then
\[ \forall i\ge2,\ e_{\ i}^{*} = A \frac{ \mu_{i} \langle e_{i}\rangle^2}{\mu_i - \mu^*} \ , \]
where $A$ can be determined by using a normalization condition for the eigenmode. 
In addition, multiplying (\ref{eq:vp_q_star_intermediaire}) by $\langle e_{i}\rangle/\left(1-\mu^{*}/\mu_{i}\right)$ and summing on $i\ge2$ yields 
\[ 1 + \frac{\mu_1 - \mu^*}{\mu_1  \langle e_{1}\rangle^2}  \sum_{i\ge2} \frac{ \mu_{i} \langle e_{i}\rangle^2}{\mu_i - \mu^*} =0 \ . \]

Finally, we obtain a condition that must satisfy $\mu^*$ and the expression of $\widetilde{e}^*$:
\begin{equation}
\sum_{i\ge 1}\frac{\mu_{i}\langle e_{i}\rangle^2}{\mu_{i}-\mu^{*}}=0\ \quad\text{with}\quad \widetilde{e}^{*}= A \sum_{i\ge 2}\frac{\mu_{i}\langle e_{i}\rangle^2}{\mu_{i}-\mu^*}\ e_{i}\ ,\label{eq:e_star}
\end{equation}
The main result is then that  that $-\mu^*$ is the smallest zero of the function\footnote{Using $1=\sum_{i\ge1}\left\langle e_{i}\right\rangle e_i$ and averaging this expression on the horizontal gives  $\sum_{i\ge1} \left\langle e_{i}\right\rangle ^{2}=1$ $\sum_{i\ge1}\left\langle e_{i}\right\rangle ^{2}=1$. Plugging this expression in (\ref{eq:f_beta}) yields  $f(\beta)=-1+\beta \sum_{i \ge 1}  \left< e_i\right>^{2}/\left(\lambda_i+\beta\right)$, which corresponds to the function (3.8) obtained by Chavanis and Sommeria \cite{ChavanisSommeria:1996_JFM_Classification}, using another method.} $f$ given by equation (\ref{eq:f_beta}). Note that the class of solutions associated with $\beta= -\mu^*$ are $\widetilde{q}=\alpha \widetilde{e}^*$, with $\alpha$ any real number. Using equation (\ref{eq:qtilde}), we find that the corresponding potential vorticity field are 
\begin{equation}
q=\frac{\Gamma}{\left< e_1\right>} e_1 +\alpha e^*\quad \text{with} \quad e^* =\widetilde{e}^* -\frac{\sum_{i\ge2} {e}^*_i \left< {e}_i\right> }{\left< e_1\right>} e_1 \label{eq:e_star_bis}
\end{equation}

We conclude that whatever the value of $\Gamma$, there is a single
solution to the variational problem (\ref{eq:canonical}) if and only if $\beta>-\min\left\{ \mu^{*},\mu_{1}^{\prime}\right\} $ (case 1. of subsection \ref{sub:method}).  When $\beta<-\min\left\{ \mu^{*},\mu_{1}^{\prime}\right\} $, there is no solution to the variational problem (case 2. of subsection \ref{sub:method}). When $\beta=-\min\left\{ \mu^{*},\mu_{1}^{\prime}\right\} $ (case 3. of subsection \ref{sub:method}), we have to consider two cases depending on the sign of $\mu^{*}-\mu_{1}^{\prime}$ in order to discuss the existence of a neutral direction:

\begin{itemize}
\item For $\beta=-\mu^{*}$, with $\mu^{*} < \mu_{1}^{\prime}$. Then a neutral direction with minimizers exists (case 3a of subsection \ref{sub:method}) if $L[\widetilde{e}^{*}]=0$. This leads to 
\[ \sum_{i\ge2}\left(\Gamma\frac{\langle e_{i}\rangle}{\langle e_{1}\rangle{}^{2}}\left(1+\frac{\beta}{\mu_{1}}\right)-\frac{\beta}{\mu_{i}}h_{i}+\frac{\langle e_{i}\rangle}{\langle e_{1}\rangle}\frac{\beta}{\mu_{1}}h_{1}\right)e_{i}^{*}=0\ .\]
By using (\ref{eq:e_star}) and a straightforward manipulation of
the previous equation, it yields the condition 
\begin{equation}
\Gamma^{*}=\mu^{*}\sum_{i\ge1}\frac{\langle e_{i}\rangle h_{i}}{\mu_{i}-\mu^{*}}\ .\label{eq:gamma_star}
\end{equation}
For $\Gamma\ne\Gamma^{*}$, there is no minimizer (case 3b of subsection \ref{sub:method}). 
\item For  $\beta=-\mu_{1}^{\prime}$, with $\mu_{1}^{\prime} <\mu^{*}$. Then a neutral direction with minimizers exists (case 3a of subsection \ref{sub:method}) if $L[e_{1}^{\prime}]=0$. This gives the condition $h_{1}^{\prime}=0$. If $h_{1}^{\prime}\ne0$, there is no minimizer
(case 3b of subsection \ref{sub:method}).
\end{itemize}
From this analysis, we conclude that three different cases have to
be considered to describe the solutions in the phase diagram $(\Gamma,E)$, depending on the sign of $\mu_{1}^{\prime}-\mu^{*}$ and on the value of $h_{1}^{\prime}$
(the projection of the topography on the smallest zero-mean Laplacian
eigenmode):
:
\begin{itemize}
\item i) $\mu^{*}<\mu_{1}^{\prime}$ or 
\item  ii) $\mu^{*}>\mu_{1}^{\prime}$
and $h_{1}^{\prime}=0$ 
\item iii) $\mu^{*}>\mu_{1}^{\prime}$ and $h_{1}^{\prime}\ne0$. 
\end{itemize}
For each of the three cases, the solutions of the canonical variational problem, as well as their energy and circulation, are computed in Appendix \ref{app:solutions_canonical}. 
In all cases we find that all admissible values of circulation $\Gamma$ and energy $E$ are reached by these canonical solutions.
We conclude the canonical problem is equivalent to the microcanonical one. 
In addition,  since there is no grand-canonical solutions above the parabola $E_{-\mu_{1}}(\Gamma)$ given by equation (\ref{eq:parabola_inequivalence}), grand-canonical  and canonical ensembles are not equivalent for this range of parameters.
The detailed computations of equilibriums states carried in  Appendix \ref{app:solutions_canonical} also lead to three different
classes of phase diagrams, corresponding to the three cases above:
\begin{itemize}
\item \textbf{case i} ($\mu^{*} < \mu_{1}^{\prime}$): there is a
single equilibrium state at each point $(\Gamma,E)$ of the phase
diagram, except on the half line $(\Gamma^{*}$, $E>E^{*})$ where
each point is associated with two states parameterized by two different
values of $\gamma.$ The parameters $\Gamma^{*}$ and $E^{*}$ are
given by equations (\ref{eq:gamma_star},\ref{eq:E_star}). At high energy, the flow structure is dominated by the eigenmode $e^{*}$
defined by (\ref{eq:e_star}) and (\ref{eq:e_star_bis}). 
\item \textbf{case ii} ($\mu^{*} > \mu_{1}^{\prime}$ and $h_{1}^{\prime}=0$):
there is a single equilibrium state at each point $(\Gamma,E)$ below
the parabola $E_{-\mu_{1}^{\prime}}(\Gamma)$ given by equation
(\ref{eq:parabolaE2eul}), and two equilibrium states at each point
$(\Gamma,E)$ above this parabola. These two equilibrium states are
parameterized by the value of their projection on $e_{1}^{\prime}$.
This value is zero on the parabola $E_{-\mu_{1}^{\prime}}(\Gamma)$,
and tends to infinity with increasing energy. High-energy states are
dominated by the eigenmode $e_{1}^{\prime}$, which is a dipole with
a neutral line along the symmetry axis of the domain. 
\item \textbf{case iii} ($\mu^{*} > \mu_{1}^{\prime}$ and $h_{1}^{\prime}\ne0$):
there is a single minimizer at each point $(\Gamma,E)$, and no peculiar
canonical transition line. At high energy, the flow structure is dominated
by the eigenmode $e_{1}^{\prime}$.
\end{itemize}
The aim of the next sections is to detail an illustrate the peculiar thermodynamic properties of the system for these three classes of phase diagrams. In particular, we will clarify the  nature of the phase transitions occurring on the half line $\Gamma^{*},\ E>E^{*}$ in case (i) phase diagrams, and on the parabola $E_{-\mu_{1}^{\prime}}(\Gamma)$ in case (ii) phase diagrams. Notice that these transition lines all stand in the ensemble inequivalence area, i.e. above the parabola $E_{-\mu_{1}}(\Gamma)$ described by equation (\ref{eq:parabola_inequivalence}).

\section{Phase diagrams in the Euler case}

For the sake of simplicity,  we restrict ourself in this section to the case of the Euler equation (with $h=0, \ R\rightarrow+\infty$ in equation (\ref{eq:qg_model_0_bord}))  in order to present the main striking features of the phase diagrams, namely the role played by the domain geometry, the changes in the flow structure associated with phase transitions, and the observation of a bicritical point. 
\begin{figure}[t]
 \includegraphics[width=1\textwidth]{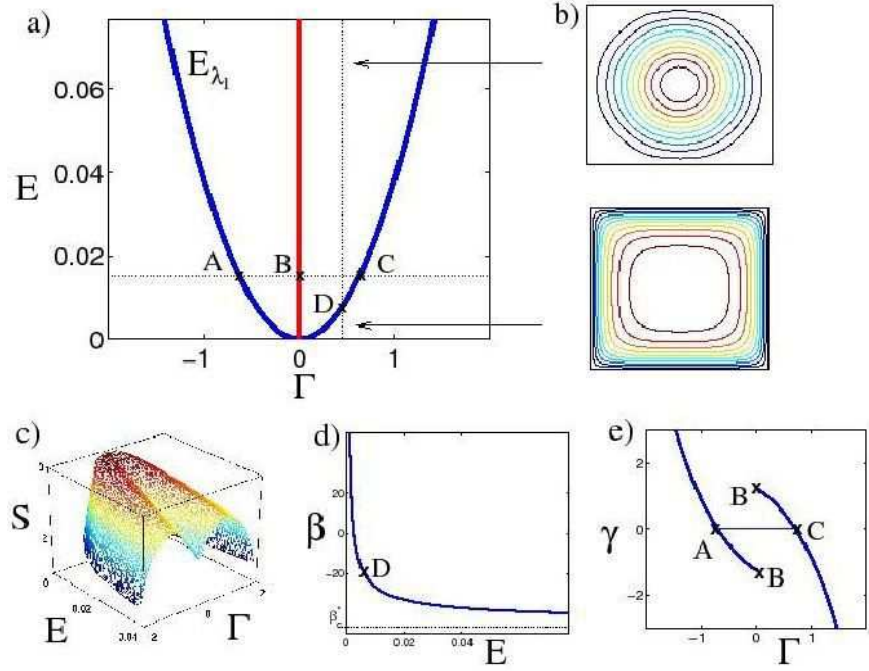}

\caption{ \textbf{ Phase diagram in case i}{ (for
a rectangular domain of aspect ratio $\tau=1.1<\tau_{c}$). Examples
of flow structures are given on panel b. The ensemble inequivalence
area is located above the blue parabola $E_{\lambda_{1}}(\Gamma)$.
The red line at $\Gamma=0$ is a first-order transition line, associated
with a positive jump of $\gamma$.}}

\label{fig:euler_ntr_phase_diag} 
\end{figure}

\begin{figure}[t]
\includegraphics[width=1\textwidth]{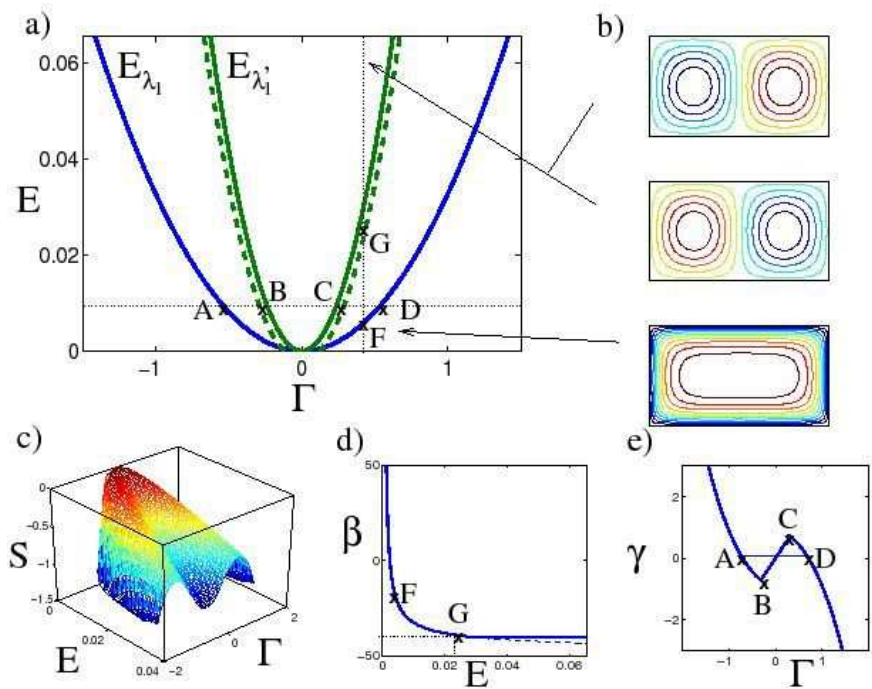}

\caption{ \textbf{ Phase diagram in case ii}{ (for a rectangular
domain of aspect ratio $\tau=1.8>\tau_{c}$). Examples of flow structures
are given on panel b. The ensemble inequivalence area is located above
the blue parabola $E_{\lambda_{1}}(\Gamma)$. The Green dashed parabola
$E_{\lambda_{1}^{\prime}}(\Gamma)$ is a second-order transition line.
Inside this parabola, $\partial^{2}S/\partial\Gamma^{2}>0$}}

\label{fig:euler_tr_phase_diag} 
\end{figure}

\subsection{A geometry governed criterion }

In the case of the Euler equations, we have $q=\Delta\psi$, so the eigenvalues $\mu_i$ defined equation (\ref{eq:def_mu}) are simply the Laplacian ones: $\mu_i=\lambda_i$. According to the previous section, there are only two types of phase diagrams when $h=0$. The criterion for one phase diagram or another depends solely on the sign of $\mu^*-\mu^\prime_1=\lambda^{*}-\lambda_{1}^{\prime}$, where $\lambda^{*}$ is the smallest zero of the function (\ref{eq:f_beta})  ($\sum_{i} \lambda_{i}\langle e_{i}\rangle / \left(\lambda_{i}-\lambda^{*}\right) =0$) and where $\lambda_1^\prime$ is the smallest eigenvalue associated with a zero-mean Laplacian eigenmode.
This criterion depends only on Laplacian eigenvalues, which depends themselves only on the domain geometry. We distinguish case (i) diagrams for which  ($\lambda^{*}<\lambda_{1}^{\prime}$) from case (ii) diagrams for which ( $\lambda^{*}>\lambda_{1}^{\prime}$).   
If the domain admits no symmetry axis, only case (i) is possible, since there is generically no zero-mean Laplacian eigenmodes in that case.
If the domain admits a symmetry axis, the sign of $\lambda_{1}^{\prime}-\lambda^{*}$ must be computed. 
In the case of a rectangular domain, there is a critical aspect ratio that can be computed numerically.
This computation gives $\tau_{c}\simeq1.12$, where the aspect ratio is defined by $\tau=L_{x}/L_{y}$ with $L_{x}$ and $L_{y}$ the lengths of the rectangular domain), and which was already reported by Chavanis and Sommeria \cite{ChavanisSommeria:1996_JFM_Classification}.
For aspect ratios smaller than $\tau_{c}$, the phase diagram is in
case (i), since $\lambda_{1}^{\prime}-\lambda^{*}>0$. 
For aspect ratios greater than $\tau_{c}$, the phase diagram is included in case (ii), since $\lambda_{1}^{\prime}-\lambda^{*}<0$.
It is expected (but not proven) that any domain geometry admitting
a symmetry axis, and sufficiently stretched in a direction perpendicular
to this axis, is in case (ii).
The figures presented in this section are computed in the case of
a rectangular domain, in which case the Laplacian eigenvalues and
eigenmodes are explicitly known (see Appendix \ref{app:rectangle}). 
According to the analysis of the previous section, these results are however generic to any domain geometry. 
Since all the two-dimensional fields ($q, \ h, \  \psi$) are decomposed on Laplacian eigenvalues in the present study, the explicit computation of these quantities (and of the associated functionals $\mathcal{E}[q]$, $\mathcal{C}[q]$ and $\mathcal{S}[q]$)  in an arbitrary geometry would require in practice a truncation of this decomposition, and the use of a numerical solver to compute Laplacian eigenmodes and eigenvalues in the given domain geometry. 

\subsection{Thermodynamic properties of the phase diagrams}

From the knowledge of the Laplacian eigenvalues, one is able to draw
the equilibrium entropy $S(E,\Gamma)$ in terms of the internal parameters
$E$ and $\Gamma$, using the computations carried out in the previous
section (the expression of the entropy in terms of the projections
$q_{i}$ is given by (\ref{eq:SS}), and the expressions of these
projections $q_{i}$ are given by equations (\ref{eq:q_continuum})
and (\ref{eq:qalpha})).
We present such plot of $S(E,\Gamma)$ on figures \ref{fig:euler_ntr_phase_diag}-c (for case (i) phase diagrams) and \ref{fig:euler_tr_phase_diag}-c (for case (ii) phase diagrams). 
All the thermodynamic properties of the phase diagrams can be deduced
from these plots.

\subsubsection*{Ensemble inequivalence area.}

The most striking feature appearing on figures \ref{fig:euler_ntr_phase_diag}-c
and \ref{fig:euler_tr_phase_diag}-c is the existence of a region
of parameters $E,\Gamma$ for which $S(E,\Gamma)$ and its concave hull do not coincide. 
Following the analysis of the previous section, this region of parameters is the one above the parabola 
\[
E_{-\lambda_{1}}(\Gamma)=\frac{1}{2\lambda_{1}\langle e_{1}\rangle}\Gamma^{2}\ ,
\]
obtained by taking $h=0$ and $\mu_1=\lambda_1$ in equation (\ref{eq:parabola_inequivalence}).
This parabola is represented as a blue line on figures \ref{fig:euler_ntr_phase_diag}-a and \ref{fig:euler_tr_phase_diag}-a, and corresponds to the upper boundary for grand-canonical solutions. 
This illustrates the result of \cite{EllisHavenTurkington:2000_Inequivalence}: there is ensemble inequivalence whenever the entropy does not coincide with its concave hull.

This is to our knowledge the first analytical description of such a generic occurrence of ensemble inequivalence in two-dimensional flows. Chavanis and Sommeria observed (using analytical computations) phase transitions in the flow structure for the Euler equations, but without relating them to the appearance of ensemble inequivalence \cite{ChavanisSommeria:1996_JFM_Classification}. Ellis, Haven and Turkington  did compute numerically a specific case of statistical equilibria showing ensemble inequivalence, but without understanding of what control the appearance of this situation, and without deciphering the link between the ensemble inequivalence and the phase transitions \cite{EllisHavenTurkington:2002_Nonlinearity_Stability}.

We describe in the following peculiar microcanonical phase transitions associated with the existence of such ensemble inequivalence.

\subsubsection*{Phase transitions for case (i) phase diagrams.}

By taking $h=0$ in equations (\ref{eq:gamma_star}) and (\ref{eq:E_star}),
we obtain $\Gamma^{*}=0,\ E^{*}=0$. The half line \[ (\Gamma=0,E>0)\]
 is represented as a thick red line in figure \ref{fig:euler_ntr_phase_diag}-a.
 We observe on panels \ref{fig:euler_ntr_phase_diag}-c and \ref{fig:euler_ntr_phase_diag}-e the occurrence of a discontinuity of $\gamma=\partial S/\partial\Gamma$ across this line (see Appendix \ref{app:compute_phase_transition} for explicit computations). 
This is the signature of a first-order phase transition.
We also note that this discontinuity corresponds to a positive jump
of $\gamma$. 
Such a thermodynamic peculiarity would not be possible
in an ensemble equivalence area: a positive jump of the derivative
of the entropy ($\partial S/\partial\Gamma$) can only occur in a
region where the entropy $S(E,\Gamma)$ and its concave envelope do
not coincide.\\

Whatever the values of the energy $E$ and of the circulation $\Gamma$,
there is a single equilibrium state, having the structure of a monopole,
as represented at low- and high-energies on figure \ref{fig:euler_ntr_phase_diag}-d,
except on the first-order transition line, represented in red on figure
\ref{fig:euler_ntr_phase_diag}-a.
When the first-order transition line is crossed, the flow structure
changes from a monopole of a given sign to a similar monopole with
the opposite sign, with coexistence of both states on the transition
line.
Notice that the high-energy states tend to a solution dominated by
eigenmode $e^{*}$ (when $\beta\rightarrow-\lambda^{*}$), given by
equations (\ref{eq:e_star}) and (\ref{eq:e_star_bis}). The structure of this eigenmode can be
computed numerically. In the case of a rectangular domain of aspect
ratio $\tau<1.12$, it is always a monopole.

\subsubsection*{Phase transitions for case (ii) phase diagrams.}

We have shown in the previous section that \emph{above} the parabola
$E_{-\lambda_{1}^{\prime}}(\Gamma)$ given by equation (\ref{eq:parabolaE2eul}) with $\forall i, \ \mu_i=\lambda_i$,
\[
E_{-\lambda_{1}^{\prime}}(\Gamma)=\left(\frac{1}{2\left(f(\lambda_{1}^{\prime})\right)^{2}}\sum_{i\ge1}\frac{\lambda_{i}\langle e_{i}\rangle^{2}}{\left(\lambda_{i}-\lambda_{1}^{\prime}\right)^{2}}\right)\Gamma^{2}\ ,\]
there are two entropy maxima (depending of the sign of their projection
on the eigenmode $e_{1}^{\prime}$) for each point $\Gamma,E$ of
the phase diagram, while it exists only one such entropy maximum for
each point of the phase diagram \emph{below} this parabola. 
This parabola
is represented as a green line on figure \ref{fig:euler_tr_phase_diag}-b.
When the parabola $E_{-\lambda_{1}^{\prime}}(\Gamma)$ is crossed,
there is a discontinuity of $\partial^{2}S/\partial\Gamma^{2}$ and
of $\partial^{2}S/\partial E^{2}$, as observed on figures \ref{fig:euler_tr_phase_diag}-c,e and as explicitly computed in Appendix \ref{app:compute_phase_transition}.
This is the signature of a second-order phase transition.
Notice the existence of positive values of $\partial^{2}S/\partial\Gamma^{2}$
in figure \ref{fig:euler_tr_phase_diag}. This last property is similar
to negative heat capacity, but for the parameter $\Gamma$ rather
than for the energy $E$. 
This corresponds to a region of parameters
$E,\Gamma$ for which the entropy $S(E,\Gamma)$ is convex: it does
not then coincide with its concave envelope.
We conclude that such a peculiarity can occur only in the ensemble inequivalence area.\\

Below the parabola $E_{-\lambda_{1}^{\prime}}(\Gamma)$, there is
a unique equilibrium state associated with each point $(E,\Gamma)$.
These low-energy states have the structure of a monopole.
Above the parabola $E_{-\lambda_{1}^{\prime}}(\Gamma)$, each point corresponds
to two equilibrium states. 
For a given circulation, the different
states above the parabola differ only by the value of their projection
on $e_{1}^{\prime}$ (a dipole). 
The choice of one state among the
two possibilities above the parabola breaks the system's symmetry.
At high energy, this contribution dominates: the flow is therefore
a dipole.
For a fixed circulation, there is thus a continuous transition from a monopole
(at low-energy) to a dipole (at high energy).

\subsubsection*{Observation of a bicritical point}

\begin{figure}
\includegraphics[width=1\textwidth]{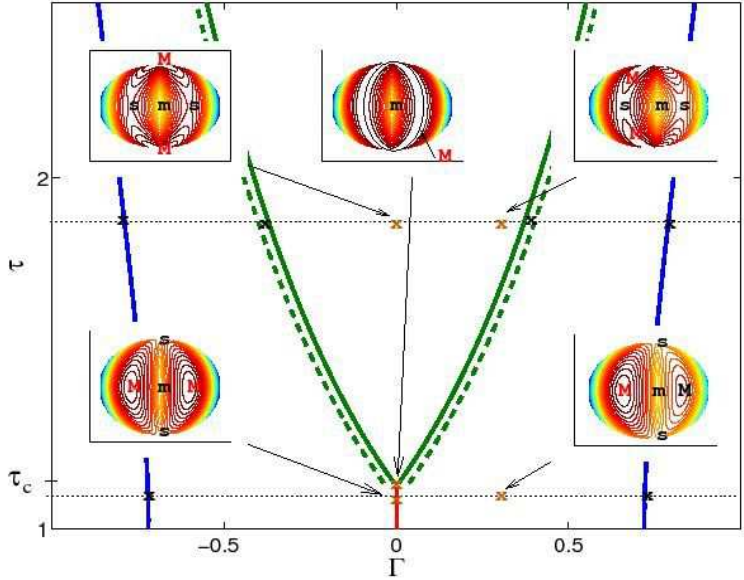}
\caption{\textbf{ Bicritical point}{: bifurcation from a first-order
phase transition line (in red) to two second-order phase transitions
lines (in green). Case of a rectangular domain, for the 2-D Euler
equation. The aspect ratio $\tau$ is taken as an external parameter,
the circulation $\Gamma$ as an internal parameter, and the energy
is fixed. Small insets are schematic representations of the entropy
$\mathcal{S}[q]$ in which the two directions correspond to projection onto the eigenmodes $e^{*}$(horizontal) and $e_{1}^{\prime}$ (vertical). }\textbf{ M}{ ,
}\textbf{ m}{ {} and }\textbf{ s}{ {} stand
respectively for }\textbf{M}{ aximum, }\textbf{ m}{ inimum
and }\textbf{ s}{ addle}}
\label{fig:bicritical_phase_diag} 
\end{figure}

Let us fix the energy and vary the aspect ratio $\tau$ in the case of a rectangular domain. Our previous analysis predicts a transition from case (i) to case (ii) phase diagrams, above a critical value $\tau_c$ of the aspect ratio. The corresponding phase diagram ($\Gamma,\tau$) is presented in figure \ref{fig:bicritical_phase_diag}. There is a bifurcation from a first-order transition line  (corresponding to the red line in case (i) phase diagram plotted in figure \ref{fig:euler_ntr_phase_diag}) to two second-order transition lines (corresponding to the green plain-dashed double line in case (ii) phase diagram plotted in figure \ref{fig:euler_tr_phase_diag}), at the point $\tau_{c}=1.12,\Gamma=0$. Such a bifurcation from a first-order to two second-order transitions is referred to as a bicritical point. 
Bicritical points have already been observed (and actually defined) in the context of short-range interacting systems, see e.g. \cite{Fisher74}. However, there is to our knowledge no example of such bifurcations for systems
with long-range interactions, while their possible existence was predicted
in \cite{Bouchet_Barre:2005_JSP}. More importantly, the previous analysis shows that this bicritical point occurs in the ensemble inequivalence area, and is in that respect a signature of ensemble inequivalence, associated with drastic changes in the flow structure (a transition from a monopole to a dipole).\\

Small insets on figure \ref{fig:bicritical_phase_diag} are schematic
representations of the functional $\mathcal{S}[q]=- \left\langle q^2 \right\rangle / 2 $ for a fixed energy and different values of the parameters $\Gamma$, $\tau$. The horizontal direction corresponds to the eigenmode $e^{*}$ defined by (\ref{eq:e_star}) and (\ref{eq:e_star_bis}) and  associated with eigenvalue $\lambda^{*}$. The vertical direction corresponds to the eigenmode $e_{1}^{\prime}$, the smallest Laplacian eigenmode with zero mean-value, associated with $\lambda_{1}^{\prime}$.
Below the green dashed line (corresponding to $\beta=-\lambda_{1}^{\prime}$)
the maxima of $\mathcal{S}[q]$ have no contribution along $e_{1}^{\prime}$, consistently with the computation of the corresponding equilibrium states, given by (\ref{eq:q_continuum}).
On the the line $\Gamma=0$, for aspect ratio $\tau<\tau_{c}$, there are
two entropy maxima proportional to $e^{*}$, with $\beta=-\lambda^{*}$. (When varying circulation, one of these two maxima becomes metastable.)
On the line $\Gamma$, for aspect ratio $\tau>\tau_{c}$, there are two entropy maxima proportional to $e_{1}^{\prime}$, with $\beta=-\lambda_{1}^{\prime}$.
The bicritical point corresponds to the transition from case (i) to case case (ii), for which $\lambda_{1}^{\prime}=\lambda^{*}$. Equilibrium states are degenerate at this point, since any linear combination of the two corresponding eigenmodes (satisfying the constraint) are entropy maxima.

\subsection{Fixed energy and varying circulation: generic occurrence of ensemble inequivalence, analogy with a magnetic system}

\begin{figure}[t]

\begin{center}
\includegraphics[width=0.7\textwidth]{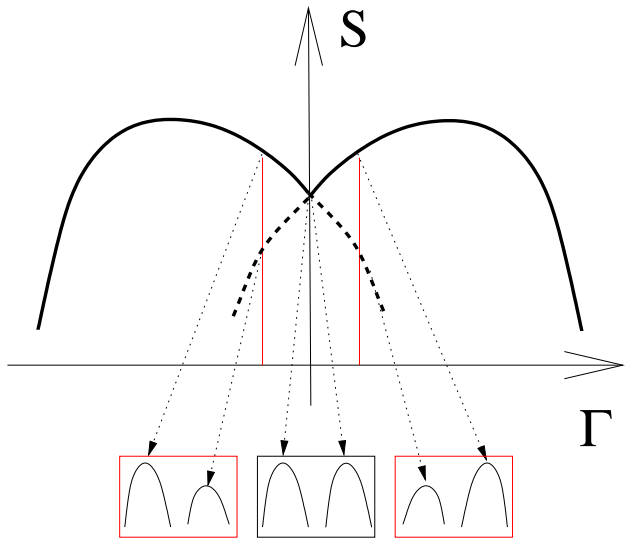}
\end{center}
\caption{Schematic representation of ensemble inequivalence, for a fixed energy and a functional $\mathcal{S}=\left<s(q)\right> $ with $s(q)=s(-q)$. Inserts represent $\mathcal{S}[q]$ for a given value of the parameter $\Gamma$. there are two symmetric solutions for $\Gamma=0$.}
\label{fig:branches}
\end{figure}
Strikingly, in all the phase diagrams described above, the ensemble inequivalence, visualized when the entropy does not coincide with its concave hull, is visible when drawing $S(E,\Gamma)$ with $E$ fixed and $\Gamma$ varied, but not visible when drawing  $S(E,\Gamma)$ with $\Gamma$ fixed and $E$ varied. This phenomenon were already mentioned in  \cite{EllisHavenTurkington:2002_Nonlinearity_Stability}. Similarly, we report positive values of $\partial^2 S/\partial \Gamma^2$, but not of  $\partial^2 S/\partial E^2$ (that would correspond in that case to negative heat capacity).\\
One can understand why ensemble inequivalence occurs  generically for the parameter $\Gamma$ in variational problems of the form (\ref{eq:microcanonical_var_prob})  when $\mathcal{C}=\langle q \rangle$, and with the symmetry  $s(q)=s(-q)$. Let us consider to simplify the case of a domain without symmetry, for a given, fixed energy.
Then the symmetry $q \rightarrow -q$ is associated with the generic occurrence of a first-order transition line for $\Gamma=0$, on the plane $E\,\Gamma$.
The reason for the occurrence of this phase transition is that they are necessarily two symmetric equilibrium states on the line $\Gamma=0$: if $q$ is a solution of the variational problem  (\ref{eq:microcanonical_var_prob}), then $-q$ is also a solution. These two states are on two branches of solutions, symmetric with respect to $\Gamma=0$, see figure \ref{fig:branches}. Varying the circulation $\Gamma$ at fixed energy $E$ favors one of the configurations (the other becoming metastable), which leads to a microcanonical first-order phase transition.
Using then the fact that  first-order phase transition in the constrained (microcanonical) ensemble  generically  implies inequivalence with the less constrained (canonical) ensemble (see figure 2 of \cite{Bouchet_Barre:2005_JSP}), we conclude that there is necessarily ensemble inequivalence for the parameter $\Gamma$ around the line $\Gamma=0$.\\

This case of ensemble inequivalence is associated with the breaking of the discrete symmetry $q\rightarrow-q$. It would similarly occur in a  magnetic system with long-range interactions, for which the computation of statistical equilibria would lead to solve the problem:

\[ S(e_{tot},\widetilde{h}) = \max_m \left\{s(m) \ | \ e_{tot}= e(m)+\widetilde{h} m \ \right \} \ ,\]
where $m$ is the magnetization, $\widetilde{h}$ is an imposed magnetic field, and where $e(m)=e(-m)$ and $s(m)=s(-m)$. Considering a fixed energy $e_{tot}$ such that the equilibrium states have a non zero magnetization when $\widetilde{h}=0$, and considering the magnetic field $\widetilde{h}$ as an external parameter, there is an ensemble inequivalence area associated with the occurrence of a first order microcanonical phase transition at $\widetilde{h}=0$. In that respect the circulation $\Gamma$ of a two-dimensional flow plays the role of an external magnetic field $\widetilde{h}$ in a magnetic system. It has to be noted however that this symmetry breaking is partial as the "field" $\Gamma$ acts only on the eigenmodes $e_i$ with $<e_i> \neq 0$.

\section{Generalization  and application to Fofonoff flows}

We explain in this section how the phase diagrams of the Euler case are changed when a non zero topography or a finite Rossby radius are taken into account. Considering the effect of the Rossby radius of deformation allows for a discussion of the behavior of the system in presence of a scale that screens the interactions. Considering the effect of topography can breaks some of the system symmetries, and allows for the description of a new peculiar thermodynamic properties, namely second-order azeotropy. It also allows for a description of Fofonoff flows in the framework of RMS statistical theory.

\subsection{Effect of the topography $h(x,y)$: symmetry breaking and second-order azeotropy}

We consider the effect of a topographic term $h$ in equation (\ref{eq:qg_model_0_bord}) by considering: 
\[q=\Delta \psi+h \ ,\]
which has two physical consequences: i) a symmetry breaking, ii) the possible presence of second-order azeotropy.

\subsubsection{First physical consequence: symmetry breaking}

First, a term $h(x,y)$  generically breaks the symmetry $q\rightarrow -q$. 
Consequently, phase diagrams in parameter space $(\Gamma,E)$ will generically be non-symmetric with respect to the axis $\Gamma=0$, unless one imposes a peculiar symmetry to $h$.
Second, even if the domain geometry admits some symmetry, the topographic term $h$ generically breaks this symmetry, unless one imposes this symmetry to $h$.
This symmetry breaking is associated with the possibility of a third class of phase diagrams, referred to as case (iii), when the domain has a symmetry axis and is sufficiently stretched in the direction perpendicular to its axis. 
We have seen previously that in case (ii) phase diagrams, i.e. when $h$ satisfies the symmetry of the domain geometry, there is a second-order phase transition line in the parameter space $(\Gamma, E)$, above which the system admits two possible equilibrium states, differing only by the sign associated with the mode $e^{\prime}_1$  (a dipole).  
In case (iii) phase diagrams, the topographic term favors one of these two degenerate equilibrium states (i.e. it selects the sign of the dipole $e^{\prime}_1$), and there is therefore no second-order transition line.
To conclude, in presence of topography, there are three classes of phase diagrams.
The criteria for these phase diagram are explicitly computed in appendix \ref{app:solutions_canonical}, as well as the corresponding thermodynamic properties.
The result depends only on the value of the projection $\left< h e_1^{\prime} \right>$ of the topography  on the  eigenmode $e^{\prime}_1$: either $h^{\prime}_1=0$ (case with symmetry), or $h^{\prime}_1 \ne 0$ (generic case without symmetry), and on the  the eigenvalues $\mu_i$ defined equation (\ref{eq:def_mu}), which, in the present case ($q=\Delta\psi+h$), are simply Laplacian eigenvalues ($\mu_i=\lambda_i$): 

\begin{description}
\item [i)] ($\lambda^{*}>\lambda_{1}^{\prime}$). 
There is a first-order microcanonical transition half line, as in the Euler case, but the minimum of this line (the point $(\Gamma^{*},E^{*})$ defined by equations (\ref{eq:gamma_star}) and (\ref{eq:E_star})), is located strictly  above  
 the parabola $E_{-\lambda_{1}}(\Gamma)$ given by equation (\ref{eq:parabola_inequivalence}). 

\item [ii)] ($\lambda_{1}^{\prime}>\lambda^{*}$ and $h_{1}^{\prime}=0$).
There is a second-order transition line as in the Euler case, namely the parabola $E_{-\lambda_{1}^{\prime}}(\Gamma)$, given by equation (\ref{eq:parabolaE2eul}). 
The curvature radius of the parabola does not depend on $h$, but the minimum of the parabola does, and it is now located strictly above $E_{-\lambda_{1}}(\Gamma)$. 
This case corresponds to figure \ref{fig:azeotropy_phase_diag}.
\item [iii)] ($\lambda_{1}^{\prime}>\lambda^{*}$ and $h_{1}^{\prime}\ne0$).
There is no microcanonical phase transition above  $E_{-\lambda_{1}}(\Gamma)$, because the topography breaks the symmetry satisfied by the domain symmetry.\\
\end{description}
Note that in these three cases, there is still ensemble inequivalence above  the parabola $E_{-\lambda_{1}}(\Gamma)$ given by equation (\ref{eq:parabola_inequivalence}). 
For $h\ne0$, the minimum of this parabola is no more located on $(\Gamma=0,E=0)$, but its curvature radius is left unchanged.\\

Note also that except its role in breaking the symmetry of the system, the topography does not induce qualitative changes in the phase diagrams described in the Euler case. 
As for the structure of the equilibrium states, the topography plays an important role for low-energy states, but high-energy states are those of the Euler dynamics. 
The energy is a quadratic positive-definite functional of the field $(q-h)$. 
Low energy states are therefore those for which $q \sim h$: the structure of the potential vorticity field (and consequently of the streamline) is entirely determined by the topography field.  
By contrast, the contribution of the topography becomes negligible for high-energy states (for which $q=\Delta\psi+h \approx \Delta \psi$), which are therefore those of the Euler dynamics. 


\subsubsection{Second physical consequence: second-order azeotropy}

\begin{figure}[t]
 \includegraphics[width=1\textwidth]{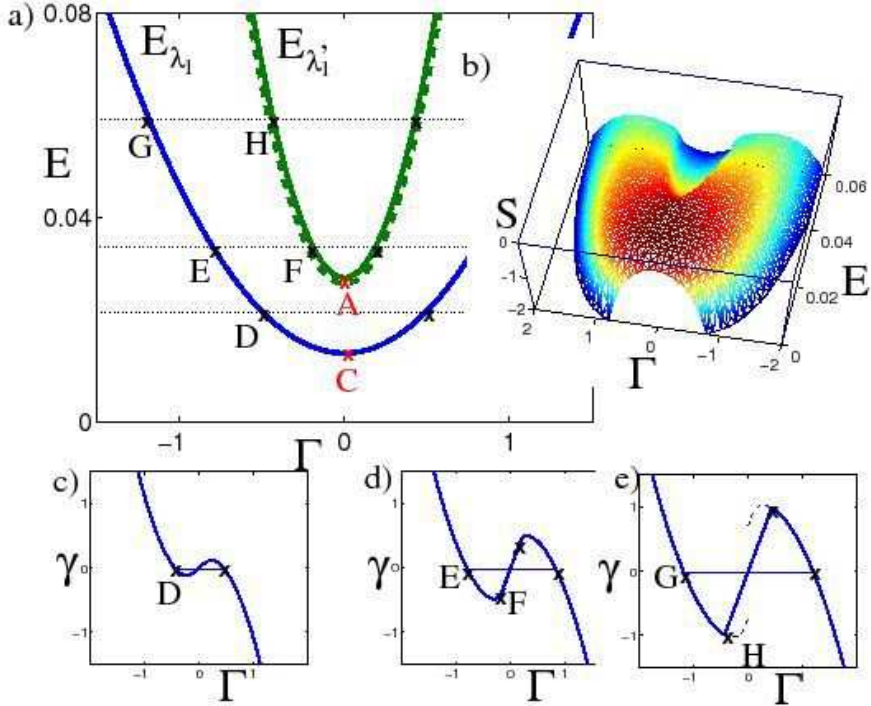}

\caption{\textbf{ Second-order azeotropy:}{ simultaneous appearance
of two second-order phase transition (at point A). Case of a rectangular
domain of aspect ratio $\tau=L_{x}/L_{y}>1.12$ with $h=\sin(\pi y/L_{y})$.}}

\label{fig:azeotropy_phase_diag} 
\end{figure}

A case (ii) phase diagram associated with a topographic term $h(y)$ is represented in figure \ref{fig:azeotropy_phase_diag}-a . Below the  parabola $E_{-\lambda_1}\left(\Gamma\right)$ represented as a blue plain line, the entropy is a concave function of the circulation, so $\gamma=\partial S/\partial \Gamma$  is a strictly decreasing function. When the  minimum $C$ of the parabola  $E_{-\lambda_1}\left(\Gamma\right)$ is crossed with increasing energy, there is the occurrence of an ensemble inequivalence area, associated with a first-order transition in the statistical ensemble with fixed energy and relaxed circulation. This first-order transition is illustrated with the Maxwell construction figure \ref{fig:azeotropy_phase_diag}-c. We conclude that point $C$ is a  critical canonical point.

When the minimum $A$ of the parabola $E_{-{\lambda_1}^\prime} \left(\Gamma\right)$ represented as a plain-dashed green line is crossed with increasing energy, there is the simultaneous appearance of two second-order phase transitions in the microcanonical ensemble. The discontinuity of $\partial \gamma/\partial \Gamma$ associated with these two second-order phase transitions can be seen in figure \ref{fig:azeotropy_phase_diag}-d, point $F$.
The simultaneous appearance of two phase transitions was referred to as second-order azeotropy in the classification \cite{Bouchet_Barre:2005_JSP}. The term \textit{azeotropic point} is commonly used for binary mixture to define the point of the phase diagram where there is the simultaneous appearance of two first-order phase transitions. This term were originally used because there is no change in the fractional composition of a binary mixture when the azeotropic point is crossed. The term ``second-order azeotropy'' extends this definition to the outbreak of two second-order phases transitions from nothing. However, contrary to the case of first-order phase transitions in binary mixtures, there is no phase separation associated with these second-order phase transitions, but rather a superpositions of one phase on the other when the transitions occur: we have seen previously that the two phases involved in the second-order transitions of case (ii) phase diagrams represent two different flow structures: a low-energy state superimposed with a ``dipole phase" whose contribution increase with the energy.

To our knowledge, second-order azeotropy has never been observed in any physical system with long-range interactions, excepted in the recent study of \cite{Stan10} in the context of the HFM model. First order azeotropy has also been reported for the first time only recently \cite{Daux10}.  As explained in \cite{Bouchet_Barre:2005_JSP}, there are two types of second-order azeotropy: one is visible in the canonical ensemble, the other is not. The azeotropic point we describe fall in this second class.
All the thermodynamic properties related to the second-order azeotropy appear as predicted in \cite{Bouchet_Barre:2005_JSP}:
\begin{itemize}
\item Sufficiently close to the azeotropic point $A$, the discontinuity of $\gamma$ is not associated with a change of concavity for the entropy: the slope of $\gamma(\Gamma)$ does not change its sign.
\item For energies sufficiently larger than the energy of the azeotropic point $A$, the discontinuity of $\partial\gamma/\partial \Gamma$  is associated with a change of concavity of the entropy: the slope of $\gamma(\Gamma)$ changes sign, as for instance seen in figure  \ref{fig:azeotropy_phase_diag}-e, point $H$.
\end{itemize}

\subsection{Application to an academic inertial ocean model, the Fofonoff flow}

\begin{figure}
\includegraphics[width=1\textwidth]{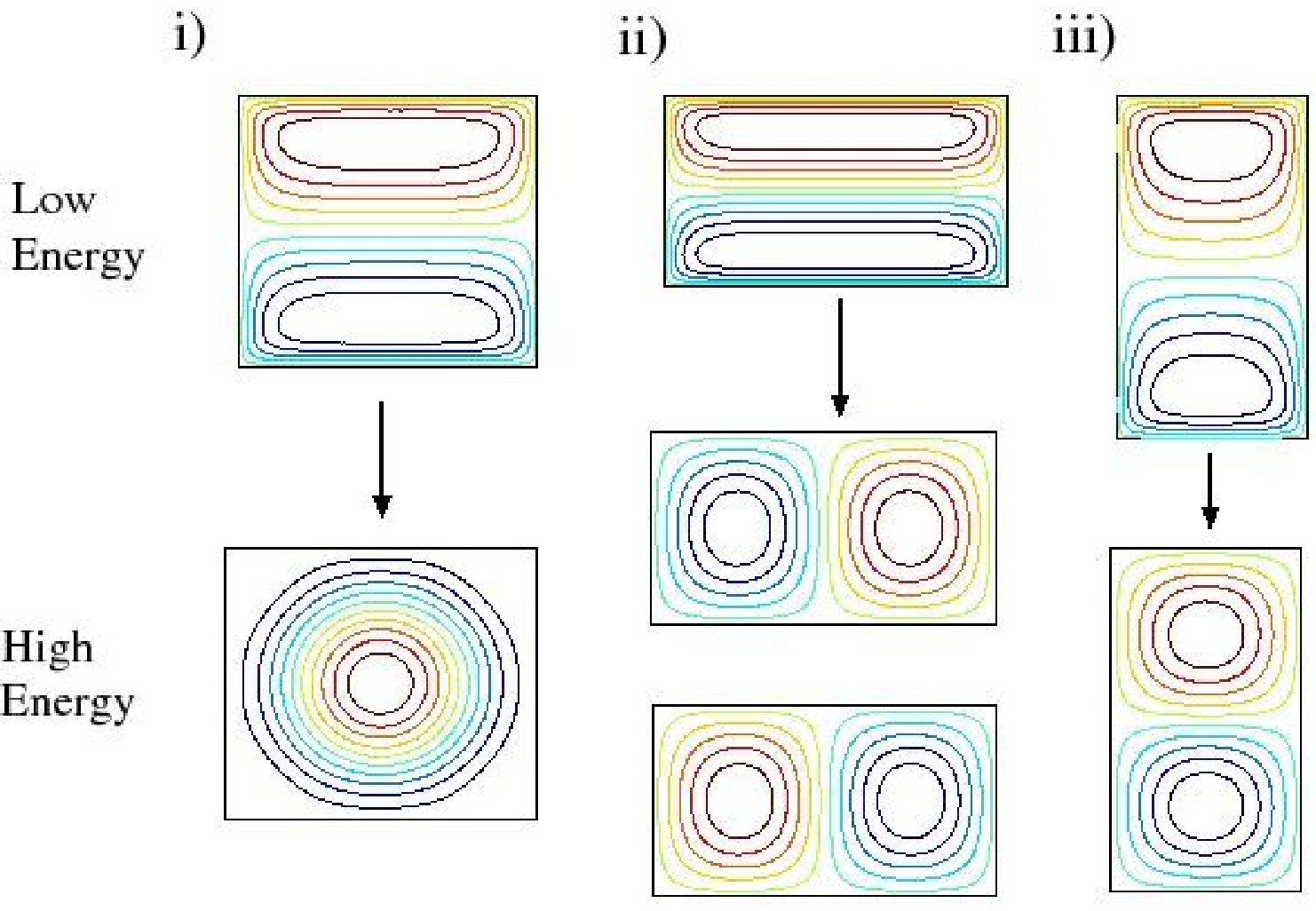}
\caption{\textbf{ Fofonoff flows, for the three possible phase diagrams
(i), (ii) and (iii)}{ The circulation is fixed, slightly greater
than zero, and $h=by$. The streamfunction is represented in a rectangular
domain with different aspect ratios (cold and hot colors are 
negative and positive values respectively). At low energy, the flow is always the
usual Fofonoff mode, with a weak westward flow in the domain bulk,
and strong recirculating jets at the boundaries. If the domain is
not stretched enough (\textbf{case (i)}), the high-energy
state is a monopole. If the domain is sufficiently stretched in the
East-West direction (\textbf{case (ii)}), there is a second-order phase transition: the high-energy state is one dipole in the
East-West direction. The choice of sign of this dipole breaks the
system symmetry. If the domain is sufficiently stretched in the North-South
direction (\textbf{case (iii)}), there is no phase transition.
The high-energy state is a dipole displaying the Fofonoff mode structure,
but without strong jets confined at the boundaries.}}

\label{fig:fofonoff_flow} 
\end{figure}

We now apply the results obtained previously to the description of an academic ocean model known as the Fofonoff solution, also referred to as Fofonoff modes or flows. Fofonoff flows are stationary solutions of the undissipated unforced barotropic quasi-geostrophic model in a closed domain $\mathcal{D}$, described by equation (\ref{eq:qg_model_0_bord}), with $h=-by$, $R=+\infty$ and $q=\beta\psi-\gamma$:

\[
\Delta\psi-\beta\psi=by-\gamma\]

 \[
\text{with}\quad\psi=0\quad\text{on}\quad\partial\mathcal{D}\ .\]

Fofonoff computed and described the solutions of these equations independently
of statistical theories, assuming $\beta\gg1$ \cite{Fofonoff:1954_steady_flow_frictionless}. The only reason to consider such an approximation was to be able to solve the mathematics. Within this approximation, the contribution of the Laplacian term is negligible in the domain bulk: 

\[ \psi\approx\frac{b}{\beta}y-\frac{\gamma}{\beta}\]
 
It corresponds to a weak westward current (the velocity in the west-east direction is $u_x=-\partial_{y} \psi \approx -b/\beta$)
taking place in the whole domain, and with strong boundary jets that close the circulation.
The emergence of such state in freely evolving numerical simulations has been previously observed and discussed by \cite{wang_vallis,ZouHolloway,Dukowicz}, and interpreted as statistical equilibrium states predicted by energy-enstrophy statistical theory in presence of topography, which leads to linear $q-\psi$ relations \cite{SalmonHollowayHendershott:1976_JFM_stat_mech_QG,bretherton}.
Energy enstrophy theory has been first proposed by Kraichnan in the framework of the statistical mechanics of truncated Euler equation \cite{Kraichnan_Motgommery_1980_Reports_Progress_Physics}. A drastic consequence of the truncation is that only the energy and the enstrophy are conserved quantities of the dynamics, while any higher vorticity moments are conserved for the Euler dynamics. This energy-enstrophy approach has then been generalized to a class of geophysical flows that include the effect of bottom topography and of the variations of the Coriolis parameter with latitude, as well as the stable stratification of these flows \cite{SalmonHollowayHendershott:1976_JFM_stat_mech_QG}. This energy-enstrophy approach has been shown to be equivalent to the phenomenological minimum enstrophy principle of \cite{bretherton}, that were proposed independently in the same geophysical context \cite{Carnevale_Frederiksen_NLstab_statmech_topog_1987JFM}. It has subsequently been shown that equilibrium states of these statistical mechanics approaches correspond to a  particular class of MRS statistical equilibria \cite{Bouchet:2008_Physica_D}, namely the class of equilibrium states computed in the present work.
Only low-energy states of barotropic ocean models on a beta plane (such as the original solution of Fofonoff) were previously computed in the framework of statistical theories. Our results make possible a direct computation of all MRS equilibrium states associated with a linear $q-\psi$ relationship model, whatever their energy and circulation. The interest is twofolds:

\begin{itemize}
\item First, it shows that the classical Fofonoff flows are MRS equilibria located in the ensemble equivalence area: there are low-energy states characterized by an inverse temperature $\beta\gg1$, for which there exists corresponding
to grand-canonical solutions. 
\item Second, we compute also high-energy states, and find that such states
can have a flow structure very different from the classical Fofonoff
solution, when they are located in the ensemble inequivalence area,
see figure \ref{fig:fofonoff_flow}. 
\end{itemize}
In addition, all the phenomenology related to phase transitions and
ensemble inequivalence applies for those models: according to the
domain geometry, one can obtain the phase diagram corresponding to
case (i), (ii) or (iii) described in the previous sections (see figure \ref{fig:fofonoff_flow}).

In particular, any domain that admits a symmetry axis in the $y$
direction, and that is sufficiently stretched perpendicularly to this
axis, corresponds to the phase diagram presented on figure \ref{fig:azeotropy_phase_diag},
for which there is second-order azeotropy.

\subsection{The effect of the Rossby radius of deformation}

We now discuss the effect of a finite value for the Rossby radius of deformation $R$ by considering:
\[q=\Delta \psi -\frac{\psi}{R^2} \ , \]
see equation (\ref{eq:qg_model_0_bord}). The only change induced by this term is that the eigenvalues $\mu_{i}$ (solutions of $\left(\Delta-R^{-2}\right)e_i=-\mu_ie_i$) are now Laplacian eigenvalues $\lambda_i$ shifted by $-1/R^{2}$.
The existence and structure of the phase diagrams (i), (ii) and (iii) are left unchanged, since the criterion   $\mu_{1}^{\prime}-\mu^{*}=\lambda_{1}^{\prime}-\lambda^{*}$ does not depend on $R$.
However, since the first Laplacian eigenvalue is of order $L^{-2}$, where $L$ is a typical length scale of the domain, taking the limit  $R \ll L$ in (\ref{eq:parabola_inequivalence}) gives $E_{-\mu_{1}}\left(\Gamma\right) \sim \Gamma^2/\left( 2 R^{-2} \right) \left<e_1\right>^2 $. 
When $R \ll L$, the curvature radius of this parabola tends to infinity.
Remembering that this parabola is the boundary for ensemble inequivalence, the range of parameters $E,\Gamma$ associated with the inequivalence between canonical and grand-canonical ensembles tends therefore  to fill the whole parameter space $E,\Gamma$.  
This result might at first sight seem surprising: the screening length scale $R$ tends to zero, which means that the energy is an additive quantity at leading order, and yet one observes an ensemble inequivalence area larger than in the Euler case !   We have explained that such inequivalence between statistical ensembles can occur for long-range interacting systems only\footnote{Ensemble inequivalence can also occur in finite volume, small short-interacting systems, see e.g. \cite{chomaz05}, but not when the thermodynamics limit is considered.}, which means that non-local terms must remain important in some way to set the structure of the equilibrium state, even if the energy is additive at leading order.                                          %

Let us be more precise: writing  $q=\zeta+\psi/R^2$, with $\zeta=\Delta \psi$, the energy reads 
\[ \mathcal{E}=-\frac{1}{2} \left(\left< \zeta \psi \right> +\frac{1}{R^2}\left< \psi^2 \right> \right) \ .\]
In the limit $R\ll L$, the term $\left< \psi^2 \right> /R^2$ dominates over the term $\left< \zeta \psi \right>$, which is the one involving long-range interaction. Similarly, the enstrophy reads
\[ -\mathcal{S}= \frac{1}{2} \left(\left< \zeta^2 \right> -\frac{2}{R^2}\left< \zeta \psi  \right>  + \frac{1}{R^4}\left< \psi^2 \right>   \right) \ , \]
and is dominated by the term  $\left< \psi^2 \right>/R^4 $ when $R \ll L$. Then, at leading order, $\mathcal{S}=-R^{-2} \mathcal{E}$, which suggests to write $\beta=-R^{-2} + \widetilde{\beta}$. Finally, the dominant term of $\mathcal{E}$ and $\mathcal{S}$ cancel each other when computing $-\mathcal{S}+\beta \mathcal{E}$. To find canonical solutions, one looks for the minimum of  
\[-\mathcal{S}+\beta \mathcal{E}= \frac{1}{2 R^2} \left( \left< \zeta \psi  \right>   +  \left< \psi^2  \right> +o\left(\frac{R^2}{L^2}\right)  \right) \ ,\]
with the constraint on the circulation. At leading order, this problem is then to find the minimum of  $\left< \zeta \psi  \right>   +  \left< \psi^2  \right>$ with the constraint on the circulation; it is independent of $R$, and involves the nonlocal term $ \left< \zeta \psi  \right> $. This shows that this non-local term, even if negligible with respect to the total energy, is crucial to compute the statistical equilibrium state.
The fact that the screening length scale plays no peculiar role in the structure of the statistical equilibria can also by seen by considering critical points of the variational problems, which satisfy $q=\beta \left(\psi -\left<\psi \right>\right)  +\Gamma$. Writing then $\widetilde{\beta}=\beta+R^{-2}$ gives 
\[\widetilde{\beta} \left(\psi -\left<\psi \right>\right) +\Gamma=\Delta \psi \ , \] 
which is just the equation satisfied by statistical equilibria of the Euler dynamics in the case of a linear $q-\psi$ relationship. An important physical consequence is also that the parameter $R$ plays no particular role in the flow structure in that case.
These results and peculiarities rely on the strong assumption of a purely linear  $q-\psi$ relation (related to a purely quadratic functional $\mathcal{S}$).
Previous analytical studies of statistical equilibria in the limit $R\ll L$, and in the case of  $\tanh$-like $q-\psi$ relationship, lead to very different flow structure and thermodynamic properties than in the linear case, with, for instance, no ensemble inequivalence, and the formation of strong jets of width $R$ in the flow structure \cite{B:2001_These}.
Such non-linear $q-\psi$ relation are the relevant one for geophysical applications, as for instance the explanation of Jovian vortices \cite{Bouchet_Sommeria:2002_JFM} or oceanic rings and jets \cite{VenailleBouchetJPO10}.  
More detailed computations and studies will therefore be necessary to fully understand the role played by the parameter $R$ when non-linearities in the $q-\psi$ relation are taken into account.
This will be the focus of a future work.%

\subsection{Generalization to a larger class of models}

\begin{figure}
\includegraphics[width=1\textwidth]{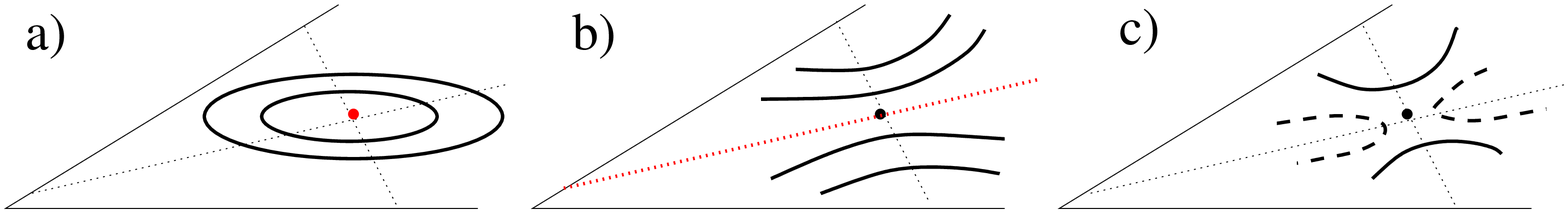}
\caption{ Schematic representation of the functional $-\mathcal{S}[q] + \beta \mathcal{E}[q]$ in configuration space, where $\beta$  is such that  \textbf{a)}  $-\mathcal{S}[q]+\beta\mathcal{E}$ is negative-definite (only stable directions);  \textbf{b)}  $-\mathcal{S}[q]+\beta\mathcal{E}$ has a neutral direction; \textbf{c)}  $-\mathcal{S}[q]+\beta\mathcal{E}$ has an unstable direction. Only case $b$ leads to non trivial solutions (the points of the neutral direction are represented as a dashed red line).}

\label{fig:generalisation} 

\end{figure}

The previous computations apply beyond the particular case of two-dimensional and geophysical flows. It can be extended to the computation of statistical equilibrium states of any system for which the energy is a quadratic functional, in the limit $E\rightarrow 0$. This could for instance be applied to the computation of statistical equilibria of the Vlasov equations. 
More generally, the results presented in the present paper hold true whenever one has to solve a variational problem that consists in maximizing  a negative-definite quadratic functional $\mathcal{I}[q]=-\mathcal{S}[q]$ of a field $q(\mathbf{r})$, with  constraints given by another definite positive quadratic functional $\mathcal{E}[q]=E$  and a linear functional $\mathcal{C}[q]=\Gamma$. 
If in addition the constraints are respectively purely quadratic (i.e. with no linear nor constant part) and purely linear (i.e. with no constant part), then there is generical ensemble inequivalence. For the sake of simplicity, let us (for instance) consider the line $\Gamma=0$ in the phase diagram ($\Gamma,E$): on this line, there are no grand-canonical solutions (i.e. no solution in the ensemble without constraints), but there are canonical solutions (i.e. solutions in the ensemble with only the circulation constraint. A schematic representation of the computation of such canonical solutions is given figure \ref{fig:generalisation}. It sums up the general method followed in this paper to compute statistical equilibria: For $\beta$ above some critical value, the (purely quadratic) functional $-\mathcal{S}+\beta\mathcal{E}$ is negative-definite, and its minimum is located at $q=0$.  For $\beta$ below some critical value, there is at least one unstable direction (one eigenmode with negative eigenvalue).  For $\beta$ equal to the critical value, there is a neutral direction, corresponding then to non trivial statistical equilibria: these are the solutions of the line $\Gamma=0$ in the phase diagram ($\Gamma,E$). 
The important conclusion is the generic occurrence of  ensemble inequivalence for this class of systems. 
Properties of phase transitions inside the ensemble inequivalence area depend on the properties of a system at hand (in particular the existence of particular symmetries).

\section{Conclusion }

The main result of this paper is the observation of a generic ensemble inequivalence region for a wide class of models, including two-dimensional and quasi-geostrophic turbulent flows, through explicit analytical computations. 
We have shown that the ensemble inequivalence is related to the occurrence of peculiar phase transitions, associated with drastic changes in the flow topology.
Several possible phase transitions have been identified in parameter space ($\Gamma, E$), leading to three different kinds of phase diagrams.
We have shown that the criterion for these different phase diagrams depends on the domain geometry only, and can be explicitly computed for a system at hand.  

Strikingly, statistical ensemble inequivalence is generically observed for varying circulation at fixed energy; we have explained that this is related to existence of a symmetry $q\leftrightarrow -q$ in the system when it admits no particular spatial symmetry. 
When varying the energy at fixed circulation, we have shown the existence of marginal cases of ensemble inequivalence at sufficiently high-energy (entropy then becomes a linear function of energy).
Taking into account non-linearity  is necessary to discuss these marginal cases.
We conjecture that $\tanh$-like relations lead to ensemble equivalence (concave entropies), while   $\sinh$-like relations could be associated with ensemble inequivalence areas. 
Understanding the role of this nonlinearity will be the focus of a future work  \cite{CorvellecBouchet2010}. 
Some of the transitions and related ensemble inequivalence we described were predicted in the classification \cite{Bouchet_Barre:2005_JSP}, but they had  not been  observed in any physical system yet.  This is the case of 
\begin{itemize}
\item Second-order azeotropy: the outbreak from nothing of two second-order phase transitions,  associated with a switch from a monopole to a dipole, breaking the system's symmetry when the energy is increased and the circulation  $\Gamma$ is kept fixed. 
\item Bicritical point: the bifurcation from a first-order transition line to two second-order transition lines when increasing the domain aspect ratio, at fixed energy, and considering the circulation $\Gamma$ as the only internal parameter. 
The first-order phase transition is associated with a switch from a monopole to another one; the two second-order transition lines correspond respectively to a switch from a monopole to a dipole, and to a switch from this dipole to another monopole, symmetric (equal and opposite) to the initial one, when varying the circulation.
\end{itemize}

Finally, we have pointed out a surprising  effect induced by a screening length scale (the Rossby radius of deformation) in 1.5 quasi-geostrophic models. Decreasing this length scale widens the ensemble inequivalence area in the phase diagrams while the interaction becomes more local. We have explained this apparent paradox, showing that non-local first-order corrections of the energy are essential to compute equilibrium states, even if the energy is additive at leading order. We have shown that this phenomenon is peculiar to the class of equilibrium states considered in this paper (states characterized by a linear $q-\psi$ relation). For such states, the flow structure does not depend on this screening length scale, which can be absorbed in the $q-\psi$ relation. We conjecture that such peculiarity would not be observed in phase diagrams obtained by considering a varying energy and fixed distributions of potential vorticity levels.\\ 
All states computed in this paper could have been observed in the framework of energy-enstrophy theories, which all lead to the same variational problem as the one we have studied in this paper. Most of the solutions to this variational problem were however not described in previous studies, carried out in the framework of energy-enstrophy theories, because the computations were only performed in the grand-canonical ensemble, where all the constraints are relaxed. 

Kraichnan noted the existence of a "condensed states" for particular values of the temperature, which correspond to the microcanonical states that we  describe in the case $\Gamma=0$ (no circulation) \cite{Kraichnan_Motgommery_1980_Reports_Progress_Physics}. He concluded that such condensed states were the most interesting ones, but that there was in average no mean flow: for all values of the parameter $E$, there were always two equiprobable symmetrical states, cancelling each other in average. Consequently to these studies, it has often been said that energy-enstrophy theories do not predict the emergence of a mean flow when no topography is taken into account. The results of the present paper clarify this issue: when performing computations in the microcanonical ensemble, we have seen that there exist (possibly multiple) equilibrium states for each value of the circulation and the energy, so that a mean flow is possible.
In the presence of topography, just as in the case of Fofonoff flows, computations in the relaxed ensemble lead to low-energy statistical equilibria, even with zero circulation. But just as in the Euler case, high-energy states could not be obtained in the most relaxed, grand-canonical ensemble. It is necessary to work in the canonical ensemble to describe them. This explains why only low-energy, Fofonoff modes, had been described by previous computations of statistical equilibrium states in the presence of topography, see e.g. \cite{SalmonHollowayHendershott:1976_JFM_stat_mech_QG,bretherton}.\\

Some of statistical equilibria were already  computed and described in the Euler case by \cite{ChavanisSommeria:1996_JFM_Classification}, in the framework of the MRS statistical theory, but using another method, and without reference to thermodynamic properties of the phase diagrams in relating to the physics of long-range interacting systems. Our contribution, following \cite{Venaille_Bouchet_PRL_2009}, has been to link these previously observed phase transitions to the generic occurrence of statistical ensemble inequivalence, using a method that can be applied to a class of systems much larger than two-dimensional and geophysical turbulent flows. Subsequently to our work, some of these results on the thermodynamic properties of phase diagrams have also been presented in \cite{NasoChavanis1} (case without topography) and \cite{NasoChavanis2} (case with topography), using the method and formalism of  \cite{ChavanisSommeria:1996_JFM_Classification}.\\

One of the current challenging problems in physics of long-range interacting
systems is to come up with experimental setups for which ensemble inequivalence
properties such as negative heat capacity could be observed. So far, no physical system has been proposed to realize such an experiment. The results presented in this paper indicate that two-dimensional flows are serious
candidates to first observe ensemble inequivalence.
Phase transitions associated with the existence of ensemble inequivalence
are predicted in the microcanonical ensemble. This ensemble can show
up if the time scale for inertial organization of the flow is much
smaller than the forcing and dissipation time scales. When such separation of time scale does not exist, it is more relevant
to include effects of forcing and dissipation. The understanding of
 phase transitions taking place at equilibrium is then a starting
point for studies out of equilibrium \cite{Bouchet_Simonnet_2009}.\\


%
%

\appendix

\section{Laplacian eigenvalues of a rectangular domain \label{app:rectangle}}

We give here the Laplacian eigenmodes ($\Delta e_{i}=-\lambda_{i}e_{i}$)
in a rectangular domain. Instead of indexing the eigenmodes of the
Laplacian by $i$, we use two indices $m,n$ corresponding to the
wave numbers in the $x$- and $y$ directions, respectively:
\[
e_{mn}=2\sin\left(\pi nx/\sqrt{\tau}\right)\sin\left(\pi my\sqrt{\tau}\right)\ ,\]
where $\tau=L_{x}/L_{y}$ and the domain is supposed to be of area
unity ($L_{y}L_{x}=1$). The corresponding eigenvalues are 
\[ \lambda_{mn}=\pi^{2}\left(\frac{n}{\tau}^{2}+\tau m^{2}\right)\ ,\]
and $\left\langle e_{mn}\right\rangle =8/\left(nm\pi^{2}\right)$
for $n$ and $m$ odd, $\left\langle e_{mn}\right\rangle =0$ for
$n$ or $m$ even. 
The zero-mean eigenmode associated with the greatest eigenvalue $-\lambda_{1}$ is
$e_{1}=e_{11}$ and the zero-mean eigenmode associated with the greatest
eigenvalue $-\lambda_{1}^{\prime}$ is $e_{1}^{\prime}=e_{1\ 2}$ (if
$\tau>1$).
 We see when  considering equation (\ref{eq:energy_infty})
that in this case, $E_{m}(\Gamma)$ is the line $E=0$ in the
plane $\left(E,\Gamma\right)$.

\section{Computation of energies and circulations of grand-canonical solutions \label{app:solution_grandcanonical}}

Let us compute the energy and the circulation of all the grand-canonical solutions, by considering separately the cases $\beta>-\mu_{1}$ and $\beta=-\mu_{1}$.
In any case, a direct computation of the critical points of (\ref{eq:J})
gives \[
\forall i\ge1,\quad\left(\mu_{i}+\beta\right)q_{i}=-\left(\gamma\mu_{i}\langle e_{i}\rangle+\beta h_{i}\right).\]

For $\beta>-\mu_{1}$ and an arbitrary value of $\gamma$, the
critical points are unique and given by

\begin{equation}
\forall i\ge1,\quad q_{i}(\beta,\gamma)=\frac{-1}{\mu_{i}+\beta}\left(\gamma\mu_{i}\langle e_{i}\rangle+\beta h_{i}\right).\label{eq:q_continuum}\end{equation}

A straightforward computation gives their circulation: 
\begin{equation}
\Gamma=\sum_{i}q_{i}\left\langle e_{i}\right\rangle =-\gamma f(\beta)-\beta\sum_{i\ge1}\frac{\langle e_{i}\rangle h_{i}}{\mu_{i}+\beta}\ ,\label{eq:cr_pt_Gamma}\end{equation}
 where \begin{equation}
f(\beta)=\sum_{i\ge1}\frac{\mu_{i}\langle e_{i}\rangle^{2}}{\mu_{i}+\beta}\ .\label{eq:f_beta_app}\end{equation}

Their energy $E=\frac{1}{2}\sum_{i}\left(q_{i}-h_{i}\right)^{2}/\mu_{i}$
can be expressed either as a function of the Lagrange multiplier $\gamma$
\begin{equation}
E_{\beta}(\gamma)=\left(\sum_{i\ge1}\frac{\mu_{i}h_{i}^{2}}{{2\left(\mu_{i}+\beta\right)}^{2}}\right)+\gamma\left(\sum_{i\ge1}\frac{\mu_{i}h_{i}\langle e_{i}\rangle}{\left(\mu_{i}+\beta\right)^{2}}\right)+\gamma^{2}\left(\sum_{i\ge1}\frac{\mu_{i}\langle e_{i}\rangle^{2}}{{2\left(\mu_{i}+\beta\right)}^{2}}\right)\ ,
\label{eq:E_g}
\end{equation}
or as a function of the circulation $\Gamma$, by using expression
(\ref{eq:cr_pt_Gamma}): 
\begin{equation}
E_{\beta}(\Gamma)=\mathcal{A}_{\beta}[h]+\mathcal{B_{\beta}}[h]\Gamma+\left(\frac{1}{2\left(f(\beta)\right)^{2}}\sum_{i\ge1}\frac{\mu_{i}\langle e_{i}\rangle^{2}}{\left(\mu_{i}+\beta\right)^{2}}\right)\Gamma^{2}\ ,
\label{eq:E_G_app}
\end{equation}
where $\mathcal{A}_{\beta}[h]=\mathcal{B}_{\beta}[h]=0$ when $h=0$,
whatever the value of $\beta$.
For $\beta=-\mu_{1}$, the solutions of the neutral direction
($\ \gamma=-h_{1}/\langle e_{1}\rangle$) are parameterized by $\alpha\in\mathbb{R}$:
\begin{equation}
q_{1}=\alpha\quad\text{and\quad}\forall i>1,\quad q_{i}(\beta,\gamma)=\frac{\mu_{1}}{\mu_{i}-\mu_{1}}\left(\frac{\langle e_{i}\rangle}{\langle e_{1}\rangle}h_{1}+h_{i}\right).\label{eq:q_canonical_neutral}
\end{equation}
By computing the circulation $\Gamma$ of these states, we find that
$\alpha=\Gamma/\langle e_{1}\rangle$: for a given circulation, there
is a single solution in the neutral direction. 
It is straightforward
to check that this solution corresponds to the solution (\ref{eq:cr_pt_Gamma}) in the limit
$\beta\rightarrow-\mu_{1}$. 
Then, the solutions of the neutral
direction are located on the parabola 
\begin{equation}
E_{-\mu_{1}}(\Gamma)=\mathcal{A}_{-\mu_{1}}[h]+\mathcal{B}_{-\mu_{1}}[h]\Gamma+\frac{1}{2\mu_{1}\langle e_{1}\rangle^{2}}\Gamma^{2}\ ,\label{eq:parabola_inequivalence}\end{equation}
obtained either by using the expression (\ref{eq:q_canonical_neutral})
or by taking the limit $\beta\rightarrow-\mu_{1}$ in (\ref{eq:E_G}).\\

We conclude that each point of the phase diagram $(E,\Gamma)$ located
on the set of parabolas $\left\{ E_{\beta}(\Gamma)\ |\ \beta>-\mu_{1}\right\} $
corresponds to a unique grand-canonical solution, where $E_{\beta}(\Gamma)$
is given by (\ref{eq:E_G_app}).

\section{Computation of energies and circulations of canonical solutions \label{app:solutions_canonical}}

The aim of this Appendix is to compute energy and circulation of the solutions of the canonical problem (\ref{eq:canonical}).
When $\beta>-\min\left\{ \mu^{*},\mu_{1}^{\prime}\right\} $, using equation (\ref{eq:cr_pt_Gamma}), one can express $\gamma$ as a function
of $\Gamma$ since $f(\beta) \ne 0$, and substitute this expression into equation (\ref{eq:q_continuum})
to obtain the expression of the solutions $q(\beta,\Gamma)$. 
The corresponding energy $E_{\beta}(\Gamma)$ is given by (\ref{eq:E_G}).
There is a single canonical solution at each point of the region spanned
by the set of parabolas $\left\{ E_{\beta}(\Gamma)\ |\ \beta>-\min\left\{ \mu_{1}^{\prime},\ \mu^{*}\right\} \right\} $.
Since $\min\left\{ \mu_{1}^{\prime},\ \mu^{*}\right\} >\mu_{1}$,
and since the curvature of these parabolas is a decreasing function
of $\beta$, we can see that the set of grand-canonical solutions
is included in the set of canonical solutions.
 Three cases must  be considered to describe the solutions in the phase diagram $(\Gamma,E)$, 
 \begin{itemize}
\item i) $\mu^{*}<\mu_{1}^{\prime}$ or 
\item  ii) $\mu^{*}>\mu_{1}^{\prime}$
and $h_{1}^{\prime}=0$ 
\item iii) $\mu^{*}>\mu_{1}^{\prime}$ and $h_{1}^{\prime}\ne0$. 
\end{itemize}

\subsubsection*{Case i ( $\mu^{*}<\mu_{1}^{\prime}$).}

There is a neutral direction for $\beta=-\mu^{*}$ if and only
if $\Gamma=\Gamma^{*}$ (where $\Gamma^{*}$ is given by equation
(\ref{eq:gamma_star})). 
In that case, $\gamma$ can not be determined by 
equation (\ref{eq:cr_pt_Gamma}). 
Without specification of the energy, $\gamma$ can then
be chosen arbitrarily on $\mathbb{R}$. By considering the expression (\ref{eq:q_continuum}) we see that  the parameter $\gamma$ parameterizes the equilibrium states of the neutral direction:
\[q_{\gamma}(-\mu^{*},\Gamma^{*})=\gamma e^{*}+\frac{\mu^{*}h_{i}}{\mu_{i}-\mu^{*}}\ ,\ \]
where $e^{*}$ is given by equations (\ref{eq:e_star}) and (\ref{eq:e_star_bis}). The energy
$E_{-\mu^{*}}\left(\gamma\ \right)$ of these states is given
by equation (\ref{eq:E_g}). 
It varies between a minimum value $E^{*}$
and $+\infty$, where 
\begin{equation}
E^{*}=\min_{\gamma\in\mathbb{R}}\left\{ E_{-\mu^{*}}\left(\gamma\ \right)\right\} \  . \label{eq:E_star}
\end{equation}
Notice that for each point $\Gamma^{*},\ E>E^{*}$, there are two
states parameterized by two different values of $\gamma$, and that
if $\Gamma\ne\Gamma^{*}$, then the energy (\ref{eq:E_G}) diverges
when $\beta\rightarrow-\mu^{*}$. 
The set of parabolas $\left\{ E_{\beta}(\Gamma)\ |\ \beta>-\mu^{*}\right\} $
hence covers the half plane $E,\Gamma$ above $E_{m}(\Gamma)$,
except the half line $\left(\Gamma=\Gamma^{*},\ E>E^{*}\right)$ which
is filled by equilibrium states of the neutral direction.
%
%



\subsubsection*{Case ii ($\mu_{1}^{\prime}<\mu^{*}$ and $h_{1}^{\prime}=0$).}

There is a neutral direction whatever the value of $\Gamma$, when
$\beta=-\mu_{1}^{\prime}$. The states of this neutral direction
are 

%
%
{} \begin{equation}
q=q(-\mu_{1}^{\prime},\Gamma)+\alpha e_{1}^{\prime},\ \alpha\in\mathbb{R}\ ,\label{eq:qalpha}\end{equation}
 where $q(-\mu_{1}^{\prime},\Gamma)$ is given by combining (\ref{eq:q_continuum})
and (\ref{eq:cr_pt_Gamma}). Their energy is \[
E=E_{-\mu_{1}^{\prime}}(\Gamma)+\frac{1}{2\mu_{1}^{\prime}}\alpha^{2}\ ,\]
 where $E_{-\mu_{1}^{\prime}}(\Gamma)$ is obtained by taking
$\beta=-\mu_{1}^{\prime}$ in (\ref{eq:E_G}):

\begin{equation}
E_{-\mu_{1}^{\prime}}(\Gamma)=\mathcal{A}_{-\mu_{1}^{\prime}}[h]+\mathcal{B}_{-\mu_{1}^{\prime}}[h]\Gamma+\left(\frac{1}{2\left(f(\mu_{1}^{\prime})\right)^{2}}\sum_{i\ge1}\frac{\mu_{i}\langle e_{i}\rangle^{2}}{\left(\mu_{i}-\mu_{1}^{\prime}\right)^{2}}\right)\Gamma^{2}\ .\label{eq:parabolaE2eul}
\end{equation}

The set of parabolas $\left\{ E_{\beta}(\Gamma)\ |\ \beta>-\mu_{1}^{\prime}\right\} $
spans the whole range of parameters $\left(E,\Gamma\right)$ located
above $E_{m}(\Gamma)$ and below $E_{\mu_{1}^{\prime}}(\Gamma)$.
The canonical solutions of the neutral direction, parameterized by
$\Gamma$ and $\alpha=q_{1}^{\prime}$ (the projection on $e_{1}^{\prime}$),
are located on the set of parabolas $\left\{ E_{\mu_{1}^{\prime}}(\Gamma)+\alpha^{2}/(2\mu_{1}^{\prime})\ |\ \alpha\in\mathbb{R}\right\} $,
which covers the half plane above $E_{\mu_{1}^{\prime}}(\Gamma)$.
%


\subsubsection*{Case iii ( $\mu_{1}^{\prime}<\mu^{*}$ and $h_{1}^{\prime}\ne0$).}

In that case, there is no neutral direction. One can easily show that
the energy of the minimizers obtained in the case $\beta>-\mu_{1}^{\prime}$
diverges when $\beta\rightarrow-\mu_{1}^{\prime}$. The set of
parabolas $\left\{ E_{\beta}(\Gamma)\ |\ \beta>-\mu_{1}^{\prime}\right\} $
covers the whole half plane above $E_{m}(\Gamma)$.

\section{Computation of phase transitions in the Euler case. \label{app:compute_phase_transition}}

\subsection{First-order phase transitions.}

Let us compute the discontinuity of the entropy across the half line
$(\Gamma=0,\ E>0)$. 
We have 
\begin{equation}
\mathcal{S}=-\frac{1}{2}\left\langle q^{2}\right\rangle =-\frac{\beta}{2}\left\langle q\psi\right\rangle +\frac{\gamma}{2}\Gamma=\beta E+\frac{\gamma}{2}\Gamma \ . \label{eq:entropie_expression1}
\end{equation}
 Let us consider the limit $\beta\rightarrow-\lambda^{*}$ for a
given value of the energy. 
The relation (\ref{eq:cr_pt_Gamma}) with 
$\mu^*=\lambda^*$ gives
\begin{equation}
\Gamma=\gamma f^{\prime}\left(-\lambda^{*}\right)\left(\beta+\lambda^{*}\right)\label{eq:Gamma_interm1}
\end{equation}
and then $\mathcal{S}=-\lambda^{*}E+\gamma\Gamma/2+\Gamma E/\left(\gamma f^{\prime}\left(-\lambda^{*}\right)\right)+O\left(\left(\beta+\lambda^{*}\right)^{2}\right)$.
Similarly, the equilibrium state (\ref{eq:q_continuum}) reads (with
first-order correction) $q\ =\gamma e^{*}+O\left(\beta+\lambda^{*}\right),$
where $e^{*}$ is given by (\ref{eq:e_star}) and (\ref{eq:e_star_bis}), with 
$\forall i,\ \mu_i=\lambda_i$. 
Then $E=-\frac{1}{2}\gamma^{2}\left\langle e^{*}\Delta^{-1}e^{*}\right\rangle +O\left(\beta+\lambda^{*}\right)$.
By noticing that $\left\langle e^{*}\Delta^{-1}e^{*}\right\rangle =f^{\prime}\left(-\lambda^{*}\right)$
and by using (\ref{eq:Gamma_interm1}), we obtain 
\begin{equation}
\gamma=\pm\kappa^{*}\sqrt{E}+O\left(\Gamma/\sqrt{E}\right)\quad\text{with}\quad\kappa^{*}=\sqrt{\frac{-2}{f^{\prime}(-\lambda^{*})}}\ ,\label{eq:gamma_interm2}
\end{equation}
which gives $q\ =\pm\kappa^{*}\sqrt{E}e^{*}+O\left(\Gamma/\sqrt{E}\right)$.
To determine the sign of the maximizer of $\mathcal{S}$ at fixed
energy and circulation, we inject (\ref{eq:gamma_interm2}) in (\ref{eq:entropie_expression1})
an use (\ref{eq:Gamma_interm1}) to find the equilibrium entropy 
\[
S=-\lambda^{*}E+\kappa_{1}^{*}\sqrt{E}|\Gamma|\ +O\left(\frac{\Gamma^{2}}{E}\right).
\]
 Thus, the sign of the maximizer $q$ is determined by the sign of
the circulation. 
When the circulation zero, both states $\pm q$ are
maximizers. 
Since $\kappa_{1}^{*}$ does not vary with $\Gamma$,
there is a positive jump of $\gamma=\partial S/\partial\Gamma$ on
$\Gamma=0$. 
We conclude that the half line $\Gamma^{*},E>E^{*}$ is a first-order
transition line. 
Two states (parameterized by two different values
of $\gamma=\partial S/\partial\Gamma$) coexists at each point of
this line, and any perturbation of the circulation select one of those
states as the unique equilibrium state. 
The entropy is a concave function
of the energy ($\partial^{2}S/\partial E^{2}<0$ for $\Gamma\ne0$),
with a marginal situation on the first-order transition line ($\partial^{2}S/\partial E^{2}=0$ for $\Gamma=0$).

\subsection{Second-order phase transitions.}

Let us compute this second-order phase transition, by considering
the limit $\beta\rightarrow-\lambda_{1}^{\prime}$ for a given energy.
We have $\mathcal{S}=\beta E+\gamma\Gamma/2$. Two cases have to be
considered : a) $\beta=-\lambda_{1}^{\prime}$ b) $\beta<-\lambda_{1}^{\prime}$.
In case a) (in which $E,\Gamma$ are located above $E_{-\lambda_{1}^{\prime}}$),
$\gamma=\Gamma/f(-\lambda_{1}^{\prime})$ and $S=-\lambda_{1}^{\prime}E+\Gamma^{2}/f(-\lambda_{1}^{\prime})$ whatever the circulation and the energy. Then $\partial^{2}S/\partial E^{2}=0$ and $\partial^{2}S/\partial\Gamma^{2}=\partial\gamma/\partial\Gamma=0$.
In case b), the relation (\ref{eq:cr_pt_Gamma}) becomes \[
\gamma=\frac{\Gamma\left(f\left(-\lambda_{1}^{\prime}\right)-f^{\prime}\left(-\lambda_{1}^{\prime}\right)\left(\beta+\lambda_{1}^{\prime}\right)\right)}{\left(f\left(-\lambda_{1}^{\prime}\right)\right)^{2}}\ +o\left(\Gamma\left(\beta+\lambda_{1}^{\prime}\right)\right).\]
%
%
The equilibrium state can be expanded at lowest order in $\beta+\lambda_{1}^{\prime}$:
\[ q\ =\gamma e_{c1}+\gamma\left(\beta+\lambda_{1}^{\prime}\right)e_{c2}\ +O\left(\Gamma\left(\beta+\lambda_{1}^{\prime}\right)\right).\]
with
\[ e_{c1}=-\sum_{i=1}^{+\infty}\frac{\lambda_{i}\langle e_{i}\rangle}{\lambda_{i}-\lambda_{1}^{\prime}}e_{i}\quad\text{and}\quad e_{c2}=\sum_{i=1}^{+\infty}\frac{\lambda_{i}\langle e_{i}\rangle}{\left(\lambda_{i}-\lambda_{1}^{\prime}\right)^{2}}e_{i}.\]
The energy reads $E=\gamma^{2}E_{c1}+\gamma^{2}\left(\beta+\lambda_{1}^{\prime}\right)E_{c2}+O\left(\beta+\lambda_{1}^{\prime}\right)$
with $E_{c1}=-\frac{1}{2}\left\langle e_{c1}\Delta^{-1}e_{c1}\right\rangle $,
$E_{c2}=-\frac{1}{2}\left\langle e_{c2}\Delta^{-1}e_{c1}\right\rangle $.  
By using the previous expression of $\gamma$, we obtain
\[ E=\frac{\Gamma^{2}}{\left(f\left(-\lambda_{1}^{\prime}\right)\right)^{2}}\left(\left(1-2f^{\prime}\left(-\lambda_{1}^{\prime}\right)\left(\beta+\lambda_{1}^{\prime}\right)\right)E_{c1}+\left(\beta+\lambda_{1}^{\prime}\right)E_{c2}\right)+O\left(\beta+\lambda_{1}^{\prime}\right)\ ,\]
which gives $\beta$ as a function of the energy and the circulation:
\[ \beta+\lambda_{1}^{\prime}=\frac{E\left(f\left(-\lambda_{1}^{\prime}\right)/\Gamma\right)^{2}-1}{E_{c2}-2E_{c1}f^{\prime}\left(-\lambda_{1}^{\prime}\right)} \ .\]
Then 
\[ \frac{\partial^{2}S}{\partial E^{2}}=\frac{\partial\beta}{\partial E}=\left(\frac{f\left(-\lambda_{1}^{\prime}\right)}{\Gamma}\right)^{2}\frac{1}{E_{c2}-2E_{c1}f^{\prime}\left(-\lambda_{1}^{\prime}\right)}\ ,\] 
and
\[\frac{\partial^{2}S}{\partial\Gamma^{2}}=\frac{\partial\gamma}{\partial\Gamma}=\frac{1}{f\left(-\lambda_{1}^{\prime}\right)}+O\left(\beta+\lambda_{1}^{\prime}\right)\ ,\] 
%
%
which is generically not equal to zero when $\Gamma\ne0$. We see that $\partial^2S/\partial E^2$ and $\partial^2S/\partial \Gamma^2$ are discontinuous when the parabola  $E_{-\lambda_{1}^{\prime}}(\Gamma)$ is crossed, i.e. when $\beta=-\lambda_1^\prime$. 
Since the second-order derivative of the entropy toward the toward the energy and the circulation are discontinuous when this parabola is crossed, we conclude that this is a second-order transition line. 

\begin{acknowledgements}

{We thank J. Barre, P.H. Chavanis, and J. Sommeria, Ana Carolina Ribeiro-Teixeira for interesting discussions,  Marianne Corvellec for providing detailed and helpful comments on the manuscript, and an anonymous reviewer for pointing to us the analogy with magnetic systems. 
for their comments on this manuscript. This work was supported by
the ANR program STATFLOW ( ANR-06-JCJC-0037-01 ) and the ANR program STATOCEAN (ANR-09-SYSC-014).}

\end{acknowledgements}

\bibliographystyle{spmpsci}      
\bibliographystyle{spphys}       
\bibliography{../../bib/Long_Range,../../bib/FBouchet,../../bib/Ocean,../../bib/Meca_Stat_Euler}

\end{document}